\documentclass[aps,prx,onecolumn,superscriptaddress,longbibliography]{revtex4-2}

\usepackage{amsmath,amsfonts,bbm}
\usepackage[colorlinks=true,breaklinks=true,linkcolor=magenta,citecolor=magenta,urlcolor=magenta]{hyperref}
\usepackage{comment}
\usepackage{graphicx}
\usepackage{floatrow}
\usepackage{mathtools}
\usepackage{booktabs}
\usepackage{diagbox}
\usepackage{listings}
\usepackage{parskip}
\usepackage{braket}
\usepackage{makecell}
\usepackage{tabularx}
\usepackage{verbatim}
\usepackage{float}
\usepackage{url}
\usepackage{subcaption}
\usepackage{algorithm}
\usepackage{algpseudocode}
\usepackage[bb=boondox]{mathalfa}
\usepackage{tikz}
\usetikzlibrary{quantikz}

\floatname{algorithm}{Procedure}

\newfloatcommand{capbtabbox}{table}[][\FBwidth]

\begin{document}

\title{Hamiltonian Simulation of Quantum Beats in Radical Pairs Undergoing Thermal Relaxation on Near-term Quantum Computers} 
\author{Meltem Tolunay}
\affiliation{IBM Quantum, 650 Harry Road, San Jose, CA 95120, USA} \affiliation{Department of Electrical Engineering, Stanford University, 350 Jane Stanford Way, Stanford, CA 94305, USA}

\author{Ieva Liepuoniute}
\affiliation{IBM Quantum, 650 Harry Road, San Jose, CA 95120, USA}

\author{Mariya Vyushkova}
\affiliation{Center for Research Computing, University of Notre Dame, 814 Flanner Hall, Notre Dame, IN 46556, USA}

\author{Barbara A. Jones}
\affiliation{IBM Quantum, 650 Harry Road, San Jose, CA 95120, USA}

\begin{abstract}
Quantum dynamics of the radical pair mechanism is a major driving force in quantum biology, materials science, and spin chemistry. The rich quantum physical underpinnings of the mechanism are determined by a coherent oscillation (quantum beats) between the singlet and triplet spin states and their interactions with the environment, which is challenging to experimentally explore and computationally simulate. In this work, we take advantage of quantum computers to simulate the Hamiltonian evolution and thermal relaxation of two radical pair systems undergoing the quantum-beat phenomena. We study radical pair systems with nontrivial hyperfine coupling interactions, namely, 9,10-octalin$^+$/$p$-terphenyl-$d_{14}$ (PTP)$^-$ and 2,3-dimethylbutane (DMB)$^+$/$p$-terphenyl-$d_{14}$(PTP)$^-$ that have one and two groups of magnetically equivalent nuclei, respectively. Thermal relaxation dynamics in these systems are simulated using three methods: Kraus channel representations, noise models on Qiskit \texttt{Aer} and the inherent qubit noise present on the near-term quantum hardware. By leveraging the inherent qubit noise, we are able to simulate noisy quantum beats in the two radical pairs better than with any classical approximation or quantum simulator. While classical simulations of paramagnetic relaxation grow errors and uncertainties as a function of time, near-term quantum computers can match the experimental data throughout its time evolution, showcasing their unique suitability and future promise in simulating open quantum systems in chemistry. 
\end{abstract}

\maketitle

\section{Introduction}
Quantum computation is an emerging and rapidly growing field with transformative potential for high-performance computing. Chemistry is considered one of the most promising applications of quantum computing, due to the inherently quantum nature of chemical systems. Spin chemistry, in particular, was recently introduced as a new promising application of quantum computing \cite{rost2020simulation, sugisaki2021quantum}. This research area has potential applications in quantum biology, artificial photosynthesis, solar materials design, organic light-emitting diodes and spin-based quantum computing \cite{wasielewski2020exploiting, forbes2019molecules, hore2020spin, wasielewski1992photoinduced, kominis2015radical}, yet it remains understudied from the quantum computing applications standpoint.

There is a long history of interdisciplinary connections between spin chemistry and quantum computation. On one hand, spin chemistry expertise has contributed significantly to the development of quantum hardware, especially to spin-based qubit implementations \cite{matsuoka2016molecular, atzori2019second, forbes2019molecules, hore2020spin, nelson2017zero, wu2018covalent, nelson2020cnot, salikhov2006time, volkov2011pulse}. On the other hand, quantum theory has been applied to spin chemistry problems such as the magnetoreception phenomenon found in birds. \cite{kominis2009quantum,cai2010quantum, gauger2011sustained,kominis2011radical, tiersch2012decoherence,hogben2012entanglement, pauls2013quantum, kritsotakis2014retrodictive, zhang2015radical, guo2017quantifying, vitalis2017quantum, mouloudakis2017quantum, kominis2020quantum, fay2020quantum}. Since electrons and qubits are both spin-1/2 systems, we aim to utilize quantum computers to simulate spin chemistry quantum dynamics that may be out of reach for classical digital computers. In our recent study, we showed that the thermal relaxation of a radical pair undergoing the quantum-beat phenomena can be efficiently simulated on a quantum computer \cite{rost2020simulation}. This inspired further quantum computing studies of more complex radical pair systems with nontrivial nuclear interactions, and the development of methods to simulate their full Hamiltonian. 

\begin{figure}[H]
\centering
\includegraphics[width=14cm]{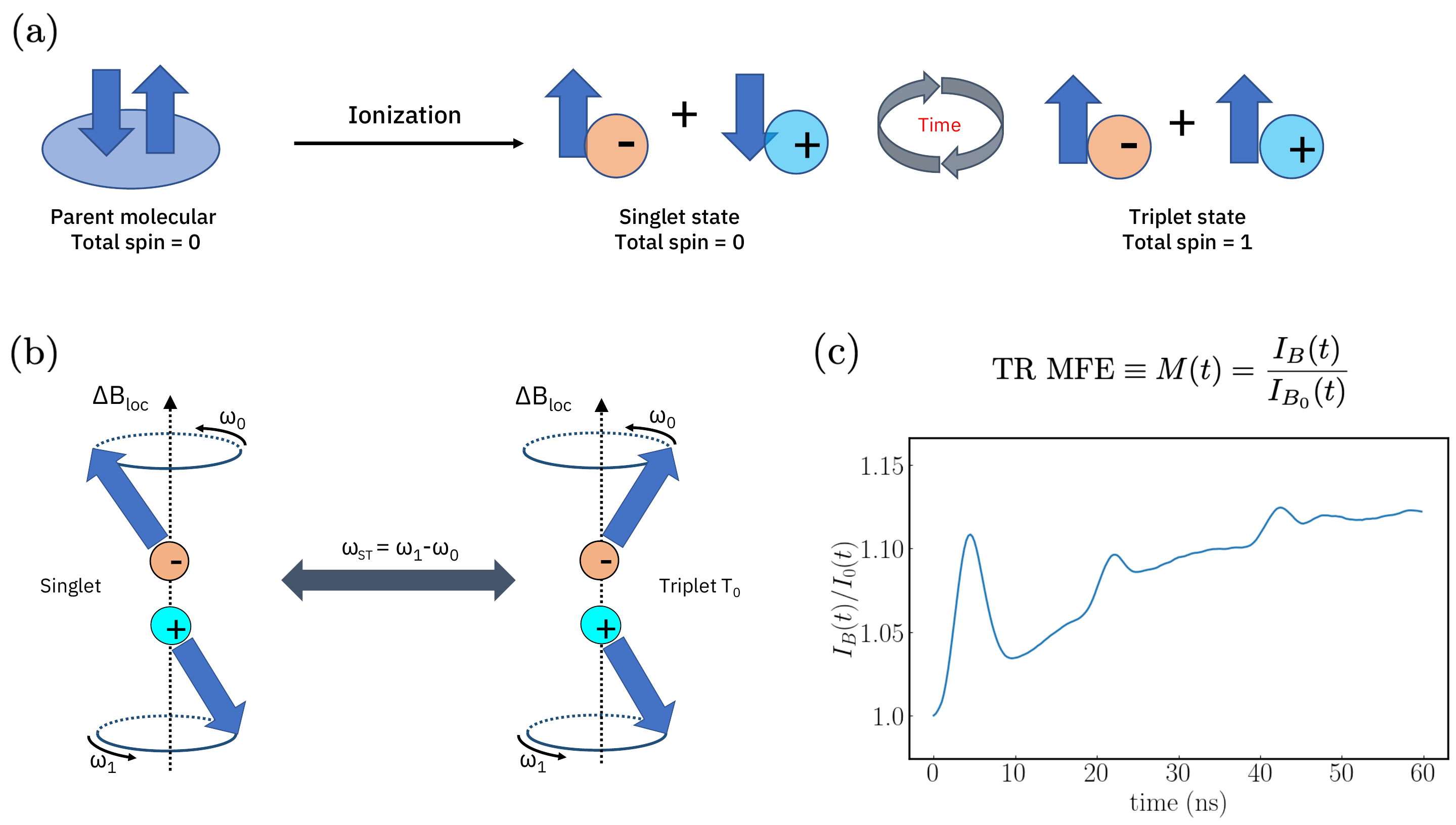}
\centering
\caption{(a) Formation and spin state evolution of an entangled singlet-triplet radical pair; (b) Vector diagram representing singlet-to-triplet oscillations in a radical pair in strong magnetic field (adapted from \cite{rost2020simulation}); (c) An example of simulated time-resolved magnetic field effect (TR MFE) plotted as a ratio of high-field and low-field recombination kinetics as a function of time for the DMB$^+$/PTP$^-$ radical pair.}
\label{fig:first}
\end{figure}

The key mechanism for magnetic and spin effects in free radical recombination kinetics is the Radical Pair Mechanism (RPM), which is based on the principle of total spin conservation in chemical reactions and singlet-to-triplet time evolution of a radical pair quantum state (Fig. \ref{fig:first}a). Entangled radical pair oscillatory behavior results in ``quantum beats'' (Fig. \ref{fig:first}b) \cite{bagryansky2007review, molin2004spin_oscillations, molin1999quantum_beats}. In a laboratory setting, a quantum beats experiment starts with a pulse of ionizing radiation passed through a dilute organic solution, which results in solvent (S) ionization as shown in equation \ref{eq:step1}. A spin-correlated radical pair is formed as a result of subsequent electron transfer reactions involving the solute molecules  \cite{bagryansky2007review, molin1999quantum_beats}. As shown in equation \ref{eq:step2}, radical cations are formed when electrons are transferred from electron donor (D) solute molecules to ionized solvent cations. At the same time, radical anions are generated when electron acceptor (A) solute molecules capture electrons from the environment (equation \ref{eq:step3}).
\begin{align}
    S &\rightarrow S^{\cdot{+}} + e^- \label{eq:step1}\\
    D + S^{\cdot{+}} &\rightarrow D^{\cdot{+}} + S \label{eq:step2}\\
    A + e^- &\rightarrow A^{\cdot-} \label{eq:step3}
\end{align}
Geminate radical pairs retain the spin state of their precursor and are found in either a singlet or triplet state. Due to hyperfine couplings and unequal Larmor precession rates under an external magnetic field,  singlet-to-triplet spin state oscillations take place that affect the spin-selective recombination product yield. Depending on the radical pair's spin state, different products are obtained:
\begin{align}
    ^{S}[A^{\cdot-} + D^{\cdot{+}}] &\rightarrow singlet\ products  \label{eq:step4}\\
    ^{T}[A^{\cdot-} + D^{\cdot{+}}] &\rightarrow triplet\ products  \label{eq:step5}
\end{align}
Oscillations of the singlet state population of radical pairs in a magnetic field are experimentally detected by measuring the recombination product fluorescence signal and its dependence on time. In each experiment, the fluorescence intensity is measured in the presence and absence of an external magnetic field, and the ratio of the two dependencies is taken. The resulting plot is referred to as a Time-Resolved Magnetic Field Effect (TR MFE) curve (Fig. \ref{fig:first}c). 

In this study, we introduce new ways of encoding spin interactions on a quantum computer and implement Hamiltonian simulation for two radical-pair systems: namely, 9,10-octalin$^+$/$p$-terphenyl-$d_{14}$(PTP)$^-$ and 2,3-dimethylbutane (DMB)$^+$/$p$-terphenyl-$d_{14}$(PTP)$^{-}$, using a quantum simulator and real quantum hardware. These radical pairs are chosen for the increasing complexity of groups of magnetically equivalent nuclei. The 9,10-octalin radical cation has a single group of magnetically equivalent nuclei. It is therefore one of the simplest type of systems with multiple nuclei exhibiting hyperfine coupling. \cite{bagryansky2000quantum}. In contrast, the DMB radical cation is a system with two groups of magnetically equivalent nuclei described by two hyperfine coupling constants and is an example of the most complex theoretical system for which there is a general analytical solution \cite{bagryansky2005spin}. We neglect the hyperfine coupling constants in PTP$^{-}$ since experimentally it is perdeuterated which makes its hyperfine coupling constants negligible. 

On quantum computers, the RPM simulation problem can be modeled as the time evolution of an entangled two-spin system with magnetic interactions (external magnetic field and/or magnetic nuclei present in the radicals). In this model, spin-correlated radical pairs are represented by electron-spin qubit pairs initialized in the maximally entangled singlet state \cite{hore2020spin, wu2018covalent, nelson2020cnot}. The spin-selective recombination of radical pairs can be treated as a quantum measurement that collapses the wave function into either the singlet or triplet eigenstate \cite{kominis2009quantum,kominis2011radical,kritsotakis2014retrodictive, jones2010spin, shushin2010effect,il2010should, ivanov2010consistent, purtov2010theory, jones2011reaction, kominis2011comment, jones2011reply, tiersch2012open, dellis2012photon, bagryansky2013verification}. 

To effectively simulate the dynamics of electron spins in radical pairs, the information of the relative orientation of the electron spin as a function of time is needed. This requires constructing an accurate Hamiltonian model for the coherent time evolution of the system while simultaneously accounting for the effects of the external environment. The two main effects governing the time dependent decay of quantum beats are paramagnetic relaxation and temperature. Classically, the effects of the paramagnetic relaxation on the singlet state population of radical pairs can be most easily described by equations that are based on semi-classical considerations only i.e. perturbation expansion where higher-order terms are neglected. While the analytic description of radical-pair oscillations at very short times without relaxation is classically possible for two or less groups of magnetically equivalent nuclei, the approximation breaks down when environmental effects and longer simulation times are considered. By utilizing quantum computers, we can take advantage of the inherent qubit noise to induce paramagnetic relaxation on long time scales. The current near-term quantum hardware has gate and qubit errors, but rather than considering this a disadvantage, we can leverage it to effectively model the time-dependent decay of radical pairs undergoing the quantum-beat phenomena. 

\section{Techniques}

\subsection{Quantum chemistry calculations}

First-principles calculations of the g-tensor and HFC constants were carried out with the ORCA 4.2.1 package \cite{neese2020orca}. The 9,10-octalin and  DMB radical cations were optimized at the unrestricted $\omega$B97xd/6-311g(d,p) \cite{chai2008long} level of theory. The minimum of the 9,10-octalin radical cation was found in a twisted chair-like geometry. The hyperfine coupling constants were calculated for all hydrogen atoms using the B3LYP functional \cite{becke98density} and the polarized triple-zeta (def2-TZVPP) basis \cite{weigend2005balanced} for carbon, and the special double zeta EPR-II (Barone’s Basis for EPR calculations) basis \cite{barone1995structure} for hydrogen atoms. All calculations were in gas phase.

\begin{figure}[H]
     \centering
     \begin{subfigure}[b]{0.45\textwidth}
     \label{fig:octalin_sub}
         \centering
         \includegraphics[width=0.5\textwidth]{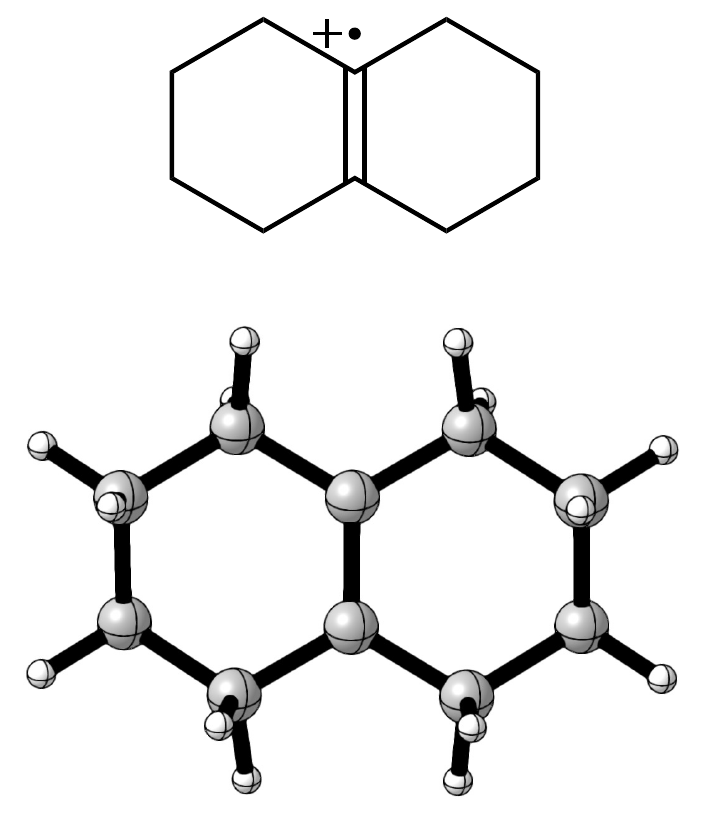}
              \caption{9,10-octalin radical cation}
     \end{subfigure}
     \hspace{-10em} 
     \begin{subfigure}[b]{0.45\textwidth}
     \label{fig:dmb_sub}
         \centering
         \includegraphics[width=0.5\textwidth]{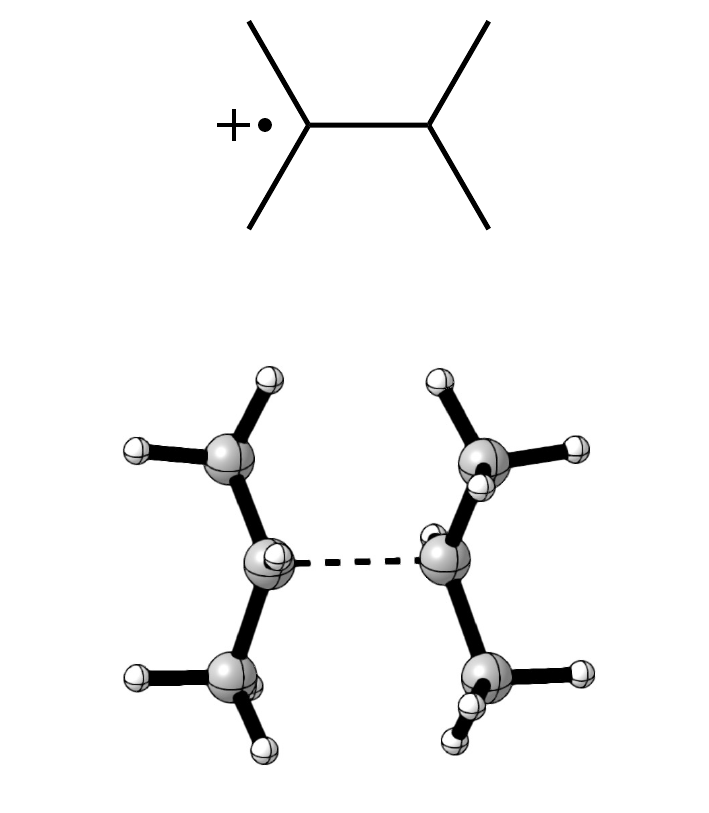}
              \caption{DMB radical cation}
     \end{subfigure}
        \caption{Structure of the (a) 9,10-octalin radical cation with a single group of magnetically equivalent nuclei and the (b) 2,3-dimethylbutane (DMB) radical cation with two groups of magnetically equivalent nuclei. The bond order between the central carbon atoms in both molecules is decreased relative to their neutral states. These radical cations are formed by the removal of an electron from the highest occupied molecular orbital (HOMO) which becomes a singly occupied molecular orbital (SOMO) resulting in a longer net bonding.}
\end{figure}

Bagryansky et al. \cite{bagryansky2005spin} performed Time-Resolved Magnetic Field Effect (TR MFE) studies of the DMB$^+$/PTP$^-$ radical pair and found that there were two possible sets of proton HFC constants $\{a_1, a_2\}$ with different signs that fit the experimental data equally well: $\{a_1\textrm{(12H)}=1.66 mT$, $a_2(2H)=0.65 mT\}$ or $\{a_1 \textrm{(12H)}=1.77 mT$, $a_2\textrm{(2H)}=-0.68 mT\}$. Since no further studies for the HFC constants of this radical cation were performed, Bagryansky et al. were unable to conclude which set of parameters was correct. Our DFT calculations allowed us to determine the correct set of experimental HFC constants, which we utilized in our quantum simulations of the spin dynamics in the DMB$^+$/PTP$^-$ radical pair.

\subsection{Basis states selection and Hamiltonian construction}

To map the Hamiltonian evolution onto a quantum computer, we need to find Hamiltonian eigenstates and the corresponding eigenenergies. The most natural choice of a basis is one which preserves all Hamiltonian symmetries and spans the Hilbert space.

\subsubsection{System with one group of magnetically equivalent nuclei: 9,10-octalin\texorpdfstring{$^+$}{+}/ PTP\texorpdfstring{$^-$}{-}radical pair}

The 9,10-octalin radical cation can be modeled as a system with eight spin-1/2 nuclear degrees of freedom and one spin-1/2 electronic subsystem. All nuclear components can be approximated to interact with the electronic subsystem with the same HFC constant $a$. Overall, the simulated Hamiltonian consists of three interactions:
\begin{align}
    H &= H_{hfc} + H_{B_1} + H_{B_2}
\end{align}
$H_{hfc}$ in equation \ref{eq:hfc} denotes the HFC interactions of the nuclei of the radical cation with its electronic subsystem, and $H_{B_1}$ and $H_{B_2}$ in equation \ref{eq:b12} contain the effect of the external magnetic field on the electronic subsystems of the radical cation and anion, respectively.
\begin{align}
 \label{eq:hfc}
    H_{hfc} = \sum_{n=1}^{8} a_n \, \textbf{I}_n \cdot \textbf{S}_1 
    &= a \, \textbf{I}_{total} \cdot \textbf{S}_1 \\
    &=  \frac{a}{2} \, \bigg[(\textbf{I}_{total} + \textbf{S}_1)^2 - \textbf{I}_{total}^2 - \textbf{S}_1^2 \bigg] 
\end{align}
\begin{align}
    H_{B_{1,2}} &= \frac{\mu_B g_{1,2}}{\hbar} \, \textbf{B} \cdot \textbf{S}_{1,2} 
    = \frac{\mu_B g_{1,2}}{\hbar} \, B^z  S_{1,2}^z
    = \frac{\mu_B g_{1,2}}{2 \hbar} B Z_{1,2}
    \label{eq:b12}
\end{align}

In equations \ref{eq:hfc} through \ref{eq:b12}, $\textbf{I}_n$ is the $n^{\textrm{th}}$ nuclear spin in the radical cation, $\textbf{S}_1$ and  $\textbf{S}_2$ are the electron spin in the radical cation and anion, respectively, $a$ is the hyperfine coupling constant in frequency units, \textbf{B} is the external magnetic field, $\mu_B$ is the Bohr magneton, $g_1$ and $g_2$ are the $g$-factor of the unpaired electron in the radical cation and anion, respectively, $B^z$ is the unidirectional magnetic field strength along the $z$-axis in units of Tesla, and $S_i^z$ is simply the Pauli-$Z$ operator divided by a constant of $2$ acting on the $i^{th}$ electron. \\

Due to spin conservation laws, the electronic subsystem of the radical pair undergoing the coherent time evolution starts in the singlet state. The nuclear subsystem with $N$ nuclei is initiated in the maximally mixed state, where all $2^N$ basis states are equally likely. In density matrix formalism, the initial state is expressed as:

\begin{align}
    \rho_0 = \frac{\mathbb{1}}{2^N} \otimes \ket{S}\bra{S}, \; \textrm{where} \; \ket{S} = \frac{\ket{\uparrow \downarrow} - \ket{\downarrow \uparrow}}{\sqrt{2}}.
    \label{eq:initial}
\end{align}

The final basis states are formed by multiplying the eight equivalent nuclear spin states by the two individual electronic spin states. In this basis, the obtained Hamiltonian is block-diagonal with block of size of at most 2×2. It is straightforward to diagonalize such Hamiltonian and obtain the eigenvalues and eigenstates. However, the initial state $\rho_0$ of the system (equation \ref{eq:initial}) is not an eigenstate. To correct for that, the electronic spins, entangled in the singlet state, need to be re-expressed in the eigenstate basis using the theory of Clebsch-Gordan coefficients \cite{ballentine2014quantum}. On the quantum computer we use a unitary transformation to achieve this. Every pure nuclear initial state becomes coupled with the initial singlet state of the electronic subsystem. We find that the singlet state couples to two and only two of the blocks of the original Hamiltonian. By identifying and enumerating the number of degenerate pure initial states, we run the Hamiltonian simulation on a very small subset of the basis states. We then classically post-process the simulation outcomes to account for the maximally mixed initial state. These observations allow us to construct and further simplify our Hamiltonian simulation circuits, the details of which are discussed in Appendix \ref{appendix:simplification}.

\subsubsection{System with two groups of magnetically equivalent nuclei: DMB\texorpdfstring{$^+$}{+}/PTP\texorpdfstring{$^-$}{-} radical pair}

Unlike the  9,10-octalin$^+$/PTP$^-$ radical pair characterized by a single HFC constant, the DMB$^+$/ PTP$^-$ radical pair comprises two different groups of magnetically equivalent nuclei. $H_{B_1}$ and $H_{B_2}$ are the same as in equation \ref{eq:b12}, while $H_{hfc}$ from equation \ref{eq:hfc} needs to be changed as follows:
\begin{align}
    H_{hfc} &= \bigg[ a_1 \sum_{n=1}^{2} \, \textbf{I}_n + a_2 \sum_{n=3}^{14} \textbf{I}_n \bigg]  \cdot \textbf{S}_1 \\
    &= \sum_{k=1}^2 \frac{a_k}{2} \, \bigg[(\textbf{I}_{total,k} + \textbf{S}_1)^2 - \textbf{I}_{total,k}^2 - \textbf{S}_1^2 \bigg] 
\end{align}
Hamiltonian solutions for a radical cation with more than one HFC constant carry additional technical challenges. In this case, the overall symmetry is the spin sum of all the electron and nuclear spins. We choose a basis consisting of the tensor product of the two groups of nuclear and two individual electronic spin states. Due to the effects of the magnetic field, the basis of individual spins is more convenient than total spin states. In this basis, the Hamiltonian is block diagonal, but the size of the blocks is larger than in 9,10-octalin$^+$ since the blocks have more self-interaction terms.

For the Hamiltonian simulation on a quantum computer, we can not simply select representative states and then build their weighted average to account for the maximally mixed nuclear initial state as is done for 9,10-octalin$^+$. This is because every nuclear initial state $\ket{I_1, m_1}\ket{I_2,m_2}$ produces a different outcome, which requires $O(N^4)$ simulations in contrast to $O(N)$ for the single-HFC-constant case. To circumvent that, we partition the Hamiltonian into several decoupled sections with high degeneracy, and make use of quantum state purification to simulate the maximally mixed initial state on each partition. In the final classical post-processing step, we combine the singlet state probability distributions to obtain the outcome for the overall maximally mixed nuclear initial state (Appendices \ref{appendix:mixed} and \ref{appendix:dmb}).

\subsection{Hamiltonian simulation}
 
To describe the time evolution of a system, we have to express its Hamiltonian as a unitary operator $U = e^{-iHt}$. Since the Hamiltonian is exponential in size, simulating quantum dynamics classically is very challenging. On a quantum computer this problem can be avoided if the unitary can be efficiently compiled into a sequence of discrete gates. On IBM's programmable quantum computers, composed of superconducting circuits, such gates are implemented with microwave pulses of different frequency and duration.  Generic Hamiltonians that act on three or more qubits can be compiled into a gate sequence consisting of local single-qubit operations and two-qubit controlled-not (CNOT) gates using the isometry decomposition method \cite{iten2016quantum} available in Qiskit \cite{Qiskit}. Additionally, we introduce Hamiltonian simplification techniques, such as KAK decomposition \cite{kraus2001optimal, vatan2004optimal, vidal2004universal}, that result in partitioned Hamiltonians acting on one or two qubits only (Appendix \ref{appendix:simplification}).

\subsection{Open system dynamics simulations}

To simulate an open quantum system we employ three different approaches, namely, quantum channel Kraus representation, Qiskit \texttt{Aer} noise models and a real quantum hardware (\textit{ibm\_lagos}) with its inherent noise profile.

\subsubsection{Kraus method}

A quantum channel can be considered as a unitary transformation acting on the total Hilbert space that includes both the system and environment ($E$) degrees of freedom \cite{wilde2013quantum}. Unitary time evolution ($U$) acting on an initial density matrix $\rho_0 \otimes \ket{0}\bra{0}_E$ that contains some pure state $\ket{0}_E$ in environment ($E$), can be represented as $U (\rho_0 \otimes \ket{0}\bra{0}_E) U^{\dagger}$. It has an equivalent representation with a set of Kraus operators $\{ K_i \}$ acting only on $\rho_0$ according to the equation:
\begin{align}
    \rho = \sum_i K_i \rho_0 K_i^{\dagger} \quad
    \textrm{such that} \quad
    \sum_i K_i^{\dagger} K_i = \mathbb{1}.
\end{align}
The two different representations of a quantum channel become equivalent when the environment is traced out using a partial trace operation as follows:
\begin{align}
    \textrm{Tr}_E \{U (\rho_0 \otimes \ket{0}\bra{0}_E) U^{\dagger} \} = \sum_i K_i \rho_0 K_i^{\dagger}
\end{align}
The quantum channel Kraus strategy entails constructing an explicit circuit for the infinite temperature amplitude damping and dephasing channels with the corresponding $T_1$ and $T_2$ parameters \cite{wilde2013quantum}. The Kraus operator representation can be implemented on a quantum computer using ancilla qubits that account for the environment. In our previous work \cite{rost2020simulation} we showed that the amplitude damping and dephasing noise channels commute with the electronic part of the Hamiltonians of interest. Since noise is present only in the electronic subsystem, we can implement the coherent time evolution and the noise channels separately. Notably, decaying both qubits at a rate $T_{1,2}$ has the same effect as decaying only one qubit at rate $T_{1,2}/2$. 

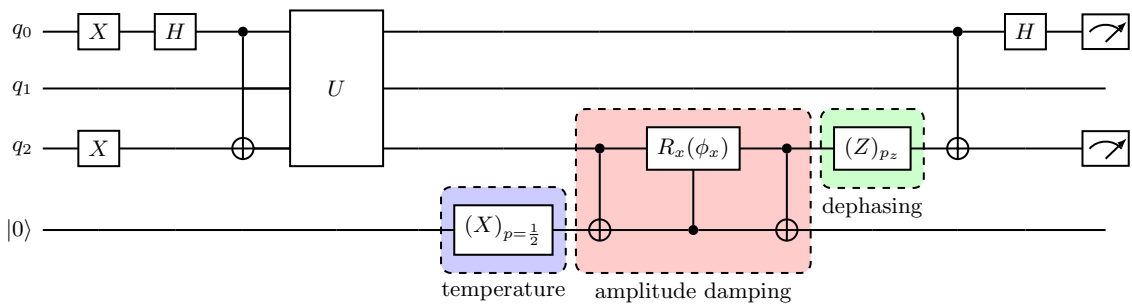
\begin{figure}[H]
    \centering
    \caption{Quantum circuit with Kraus method implementation using probabilistic gates (indicated with parentheses). Qubits $q_0$ and $q_2$ represent the electronic subsystem and $q_1$ stands for an exemplary nuclear subsystem. The last qubit is appended as an ancilla. The singlet state on the electronic qubits is initiated by applying the Pauli-$X$ ($X$), Hadamard ($H$) and CNOT gates. The unitary $U$ represents the coherent time evolution of the spin system. For thermal relaxation simulations, the corresponding parameters as functions of $T_1$ and $T_2$ are $p_x = 1 - e^{-\frac{t}{T_1}}$, $p_z = \frac{1}{2} \bigg[ 1 - e^{-t \big( \frac{1}{T_2} - \frac{1}{2T_1} \big)} \bigg]$ and $\phi_x = 2 \sin^{-1} (\sqrt{p_x})$ \cite{rost2020simulation}. $X$ gate is applied with probability $1/2$ and $Z$ gate with probability $p_z$.}
    \label{fig:circ8}
    \resizebox{0.85\textwidth}{!}{%
    \begin{quantikz}
    \lstick{$q_0$}& \gate{X} & \gate{H} & \ctrl{2} & \gate[3]{U}[2cm] & \qw & \qw & \qw & \qw & \qw & \qw & \ctrl{2} & \gate{H} & \meter{}\\
    \lstick{$q_1$}& \qw  & \qw & \qw & \qw & \qw  & \qw & \qw & \qw & \qw & \qw & \qw & \qw & \qw\\
    \lstick{$q_2$}& \gate{X} & \qw & \targ{} & \qw & \qw & \qw & \ctrl{1}\gategroup[wires=2,steps=3,style={dashed, rounded corners,fill=red!20, inner xsep=2pt}, background, label style={label position=below,anchor= north,yshift=-0.2cm}]{amplitude damping} & \gate{R_x(\phi_x)} & \ctrl{1} & \gate{(Z)_{p_z}}\gategroup[wires=1,steps=1,style={dashed, rounded corners,fill=green!20, inner xsep=2pt}, background, label style={label position=below,anchor= north,yshift=-0.2cm}]{dephasing} & \targ{} & \qw & \meter{} \\
    \lstick{$\ket{0}$}& \qw  & \qw & \qw & \qw & \qw & \gate{(X)_{p=\frac{1}{2}}}\gategroup[wires=1,steps=1,style={dashed, rounded corners,fill=blue!20, inner xsep=2pt}, background, label style={label position=below,anchor= north,yshift=-0.2cm}]{temperature} & \targ{} & \ctrl{-1} & \targ{} & \qw & \qw & \qw & \qw
    \end{quantikz}
    }%
\end{figure}

There are different quantum circuits that can implement the infinite-temperature amplitude damping and dephasing Kraus operators. The circuit implementation in Fig. \ref{fig:circ8} makes use of the fewest number of ancilla qubits \cite{rost2020simulation}. In this implementation, the two CNOT gates and the controlled-$X$ rotation account for the standard amplitude damping channel that decays $\ket{\uparrow}$ to $\ket{\downarrow}$. The probabilistic $X$ gate on the ancilla qubit that is present in the circuit with probability $1/2$ converts the zero-temperature amplitude damping channel to infinite-temperature. This is achieved by changing the direction of the amplitude decay from $\ket{\uparrow}\rightarrow\ket{\downarrow}$ to $\ket{\downarrow}\rightarrow\ket{\uparrow}$ in half of the circuit runs. 
We compute the weighted average of all configurations of the probabilistic gates for the total thermal relaxation simulation with the desired $T_1$ and $T_2$ parameters.

\subsubsection{Noise simulations with Qiskit \texttt{Aer} }

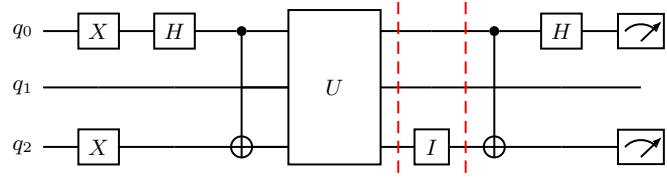
\begin{figure}[H]
    \centering
    \caption{Quantum circuit with thermal relaxation error noise model implementation as available in Qiskit \texttt{Aer}. Noise is assigned to the identity gate $I$ by specifying the desired $T_1$ and $T_2$, the gate time and the excited state population at equilibrium.}
    \label{fig:circ9}
    \resizebox{0.5\textwidth}{!}{%
    \begin{quantikz}
    \lstick{$q_0$}& \gate{X} & \gate{H} & \ctrl{2} & \gate[3]{U}[2cm]\slice{} & \qw\slice{} & \ctrl{2} & \gate{H} & \meter{} \\
    \lstick{$q_1$}& \qw & \qw & \qw & \qw & \qw & \qw & \qw & \qw\\
    \lstick{$q_2$}& \gate{X} & \qw & \targ{} & \qw & \gate{I} & \targ{} & \qw  & \meter{}
    \end{quantikz}
    }%
\end{figure}

In our second strategy to simulate an open quantum system, we use the thermal relaxation error noise model as available in Qiskit \texttt{Aer}. This model can assign noise to individual gates in a circuit, and is parameterized by relaxation time constants $T_1$ and $T_2$, the gate duration and the excited state population at equilibrium. In this approach, we insert a noisy identity gate into the circuit as shown in Fig. \ref{fig:circ9}, which is customized with the desired $T_1$ and $T_2$ relaxation parameters. The gate duration is adjusted to match the time point for which the simulation is run. The excited state population ratio at the equilibrium is set to 1/2 to account for the infinite temperature effects.

\subsubsection{Inherent qubit noise from quantum hardware}

\begin{figure}[H]
    \centering
    \caption{Quantum circuit with echo pulses implementing the inherent qubit noise method. $N$ denotes the total number of identity gates.}
    \label{fig:circ10}
    \resizebox{0.85\textwidth}{!}{%
    \begin{quantikz}
    \lstick{$q_0$}& \gate{X} & \gate{H} & \ctrl{1}\slice{} & \gate{I^{\frac{N}{8}}}\slice{} & \gate{X}\slice{} & \gate{I^{\frac{N}{4}}}\slice{} & \gate{X}\slice{} & \gate{I^{\frac{N}{4}}}\slice{} & \gate{X}\slice{} & \gate{I^{\frac{N}{4}}}\slice{} & \gate{X}\slice{} & \gate{I^{\frac{N}{8}}}\slice{} & \ctrl{1} & \gate{H} & \meter{} \\
    \lstick{$q_1$}& \gate{X} & \qw & \targ{} & \gate{I^{\frac{N}{8}}} & \gate{X} & \gate{I^{\frac{N}{4}}} & \gate{X} & \gate{I^{\frac{N}{4}}} & \gate{X} & \gate{I^{\frac{N}{4}}} & \gate{X} & \gate{I^{\frac{N}{8}}} & \targ{} & \qw  & \meter{}
    \end{quantikz}
    }%
\end{figure}
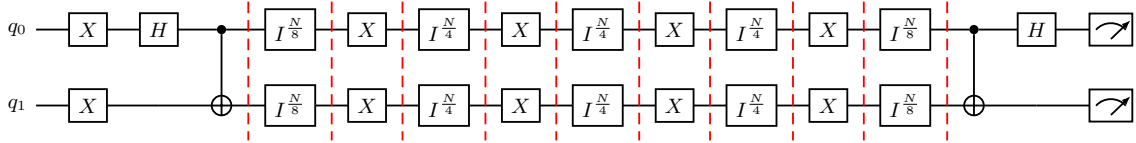

The third strategy involves simulation of the thermal relaxation of the electronic subsystem of the radical ion pairs by utilizing the inherent noise present in real quantum hardware \cite{rost2020simulation}. In this method, we classically combine several different measurement results to match the $T_1$ and $T_2$ values of the radical ion pairs to those of the quantum hardware. In general, the radical ion pairs decay with $T_1$ and $T_2$ values on the order of nanoseconds, whereas the decay constants for the IBM quantum hardware are typically on the order of microseconds. To correct for that we run additional circuits which measure the decay of the quantum hardware with the use of multiple identity gates that introduce artificial delay (Fig. \ref{fig:circ10}). Additionally, an even number of $X$ gates, referred to as ``echo pulses'', are inserted in between the identity gates \cite{rost2020simulation}. Echo pulses serve as an error mitigation technique for a hardware calibration drift that introduces phase accumulation over long simulation times. Echo pulses flip $\ket{\downarrow}$ to $\ket{\uparrow}$ during half of the simulation run-time allowing the undesired phase accumulation to cancel (Appendix \ref{appendix:inherent}).

\subsection{Classical post-processing}

To compare our quantum simulation results with the experimental data, we perform classical data post-processing that accounts for the device characteristics from the wet lab experiments \cite{bagryansky2005spin}. The rate of radical pair recombination in the singlet state is experimentally evaluated by measuring the fluorescence signal produced by the charge recombination of radiation-generated radical pairs. Fluorescence intensities $I(t)$ as a function of the singlet state probability $S(t)$ of the geminate radical pairs, are recorded at high and zero magnetic fields and reported as a ratio of the two $I_B(t) / I_0(t)$. 

In order to compare the simulated high-field $S_B(t)$ and zero-field $S_0(t)$, we let $F(t)$ denote the geminate radical pair lifetime approximated  as $F(t) = 1 / (t+t_0)^{\frac{3}{2}}$ \cite{bagryansky2005spin}, where $t_0$ depends on the chemical environment. If the experimental equipment is assumed to be ideal and the fluorescence time is ignored, the intensities at high and zero magnetic fields are captured by the equation:
\begin{align}
    \tilde{I}_{B,0}(t) &= F(t) \bigg[ \theta S_{B,0}(t) + \frac{1}{4} (1 - \theta) \bigg]
\end{align}
where $\theta$ denotes the experimentally determined geminate pair recombination rate, and $S_{B,0}(t)$ is the input from our quantum simulations. However, in an experiment, a nonzero fluorescence time is observed, and the recording device has a finite setup response time. To account for these effects, we first define $\tau_f$, the fluorescence time, as a measure of how long it takes for fluorescence to decay by a factor of $1/e$. This decay is captured by the expression $E(t) = \exp ( -t / \tau_f )$. The finite experimental setup response time  \cite{bagryansky2005spin} can be approximated as follows:
\begin{align}
    &G(t) = 
    \begin{cases}
    \frac{1}{t_g},& -\frac{t_g}{2} \leq t \leq \frac{t_g}{2}\\
    0,              & \text{otherwise} 
    \end{cases}
\end{align}
Since the mentioned effects obey linearity and time-invariance, the observed intensities can be expressed as the convolution sum:
\begin{align}
    I_{B,0}(t) &= E(t) * \tilde{I}_{B,0}(t) * G(t),
\end{align}
resulting in the observed ratio:
\begin{align}
    R = \frac{I_B(t)}{I_0(t)} = \frac{E(t) * \tilde{I}_B(t) * G(t)}{E(t) * \tilde{I}_0(t) * G(t)}.
\end{align}

\section{Results}

\subsection{Hyperfine coupling constant calculations}

Quantum chemistry calculations were performed to optimize molecular geometries and obtain a good estimate of the HFC constants for the 9,10-octalin and the DMB radical cations. Computed hyperfine couplings for each hydrogen atom in both radical cations are shown in Table \ref{tab:orca_octalin} for 9,10-octalin$^+$ and Table \ref{tab:orca_dmb} for DMB$^+$ in Appendix \ref{appendix:hfc}. Due to different electronic environments found in the ``double half-chair'' 9,10-octalin radical cation, three different sets of HFC constants are observed: ca. $0.01mT$ ($\gamma$-protons), $1.62mT$ (axial $\beta$-protons) and $4.72mT$ (equatorial $\beta$-protons). In experiment, the $\beta$-equatorial and $\beta$-axial hydrogens are averaged out due to intramolecular ring-flips with an isotropic average in good agreement to experimental values. The calculated electronic barrier for axial and equatorial hydrogen flip in 9,10-octalin is ca. $4.4 kcal/mol$, which renders axial and equatorial protons indistinguishable among half-chair ring conformations at room temperature, and results in averaging-out of the differences between the equatorial and the axial hydrogen HFCs. Experimentally, the HFC of $2.5mT$ to only one group of 8 equivalent protons is observed {\cite{EPR_QB}}. 
\begin{figure}[H]
     \centering
     \begin{subfigure}[b]{0.5\textwidth}
         \centering
         \includegraphics[width=\textwidth]{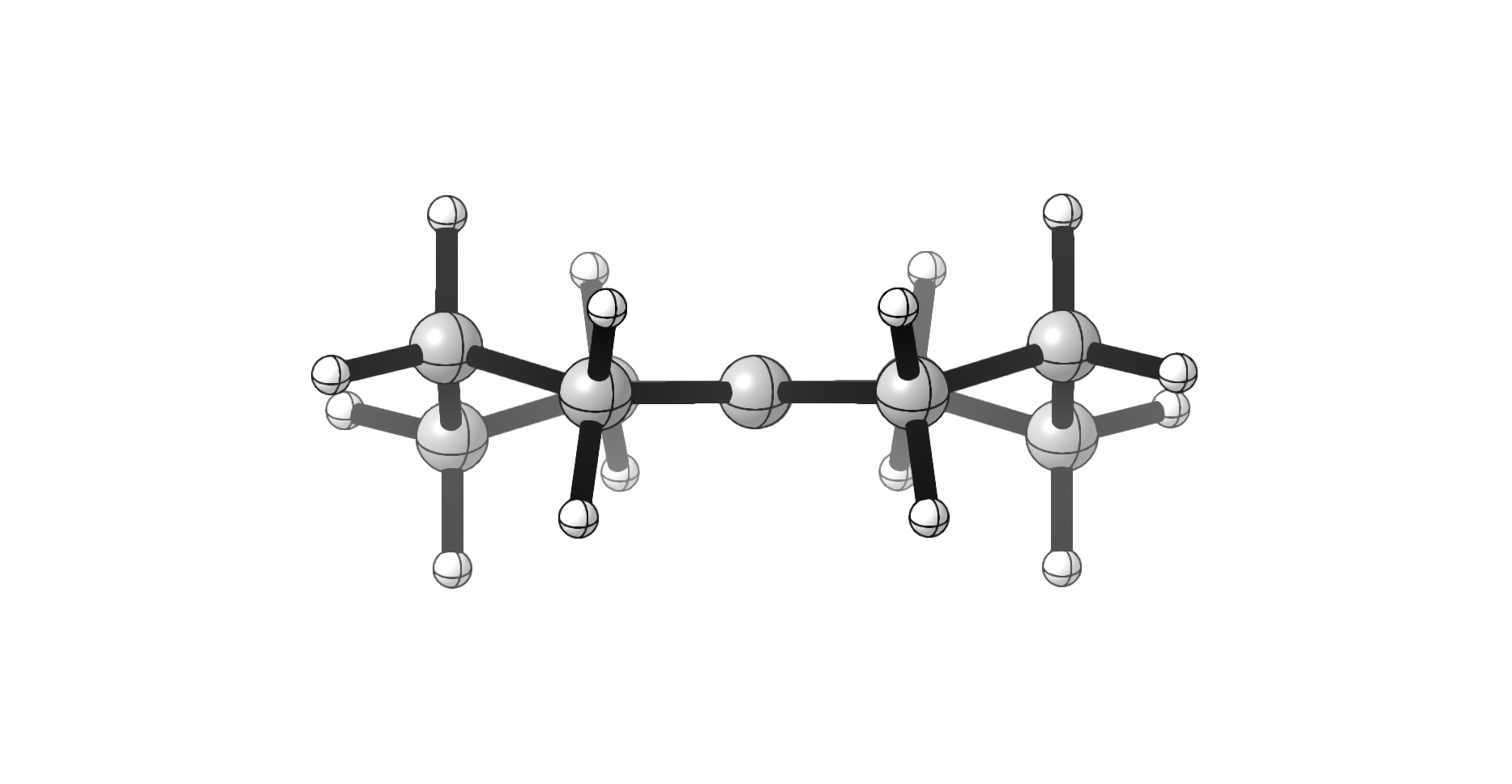}
         \caption*{$E_{rel}=0.0$}
     \end{subfigure}
     \hfill
     \begin{subfigure}[b]{0.49\textwidth}
         \centering
         \includegraphics[width=\textwidth]{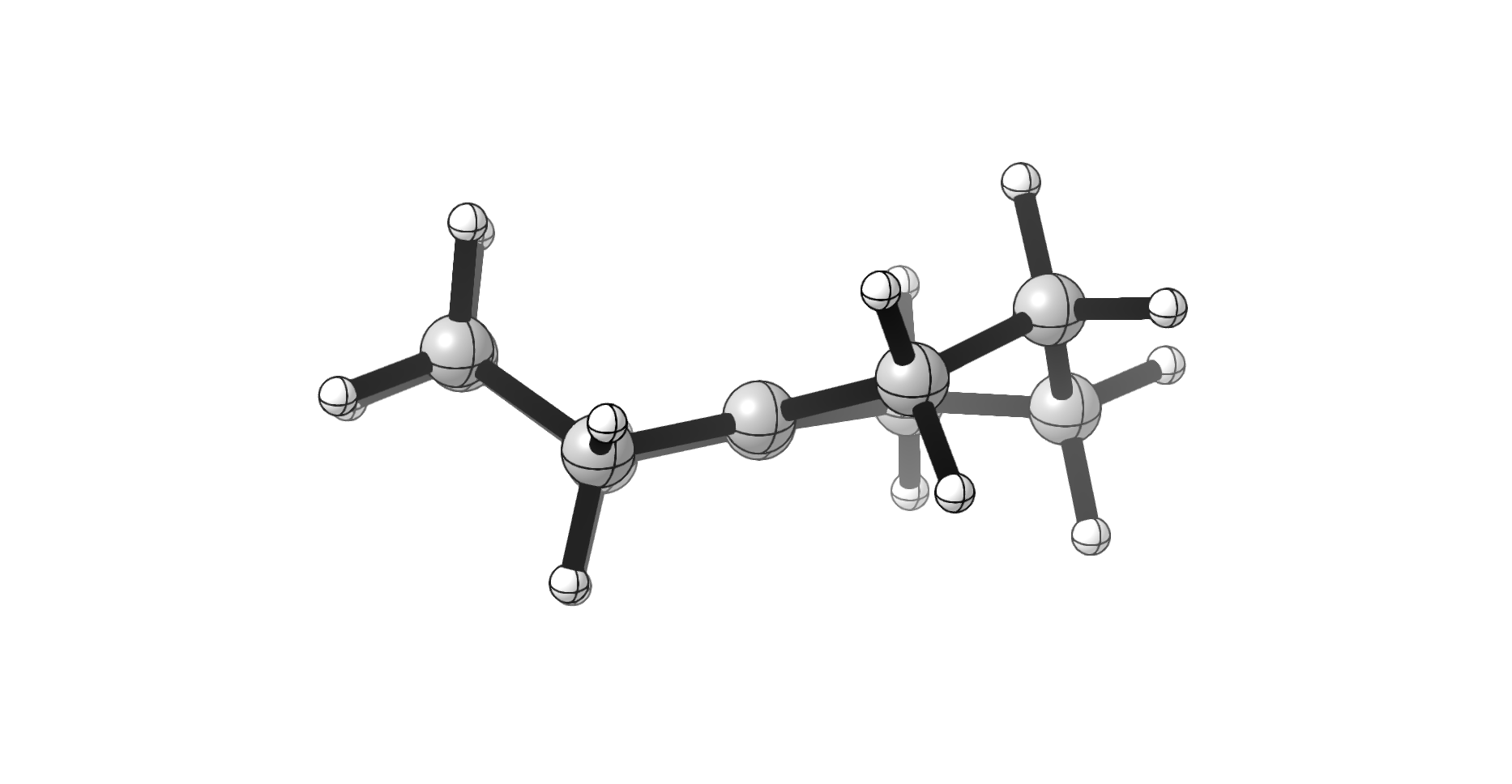}
         \caption*{$E_{rel}=4.4$}
     \end{subfigure}
        \caption{Optimized ground state geometries of 9,10-octalin (left) and its twisted configuration (right). The relative electronic energies (in kcal/mol) represent the ring flip barrier rendering axial and equatorial protons indistinguishable. Since no imaginary frequencies were found for the twisted geometry, we conclude that $4.4 kcal/mol$ is the least amount of energy needed to flip between axial and equatorial protons and more energy would be required for the ring flip transition state.}
\end{figure}

For the DMB radical cation, we compared our calculations to the experimental Time-Resolved Magnetic Field Effect (TR MFE) results obtained by 
Bagryansky et al. in an $n$-hexane solution of $0.1M$ of DMB and $30\mu M$ of $p$-terphenyl-$d_{14}$ (PTP) \cite{bagryansky2005spin}. The HFC constants of the individual hydrogens in the DMB radical cation span a wide range of values (Table \ref{tab:orca_dmb}). Interestingly, the experimental TR MFE is equally well fitted by two different sets of proton HFC constants with very close absolute values, but different signs. Since no additional experimental EPR data is available to determine which set of values is correct, our quantum chemical calculations serve as first evidence in support of the positive set of HFCs with $a_1\textrm{(12H)} = 1.66mT$ and $a_2\textrm{(2H)} = 0.65mT$. These experimental HFC constants are in good agreement with the computed values of $1.23mT$ and $0.12mT$, respectively. Slight deviations from the experimental values can be attributed to the fact that experiments were carried out in solution while our calculations were performed in the gas phase. 

\subsection{System with one group of magnetically equivalent nuclei: 9,10-octalin\texorpdfstring{$^+$}{+}/PTP\texorpdfstring{$^-$}{-} radical pair}

\begin{figure}[H]
\centering
\includegraphics[width=\textwidth]{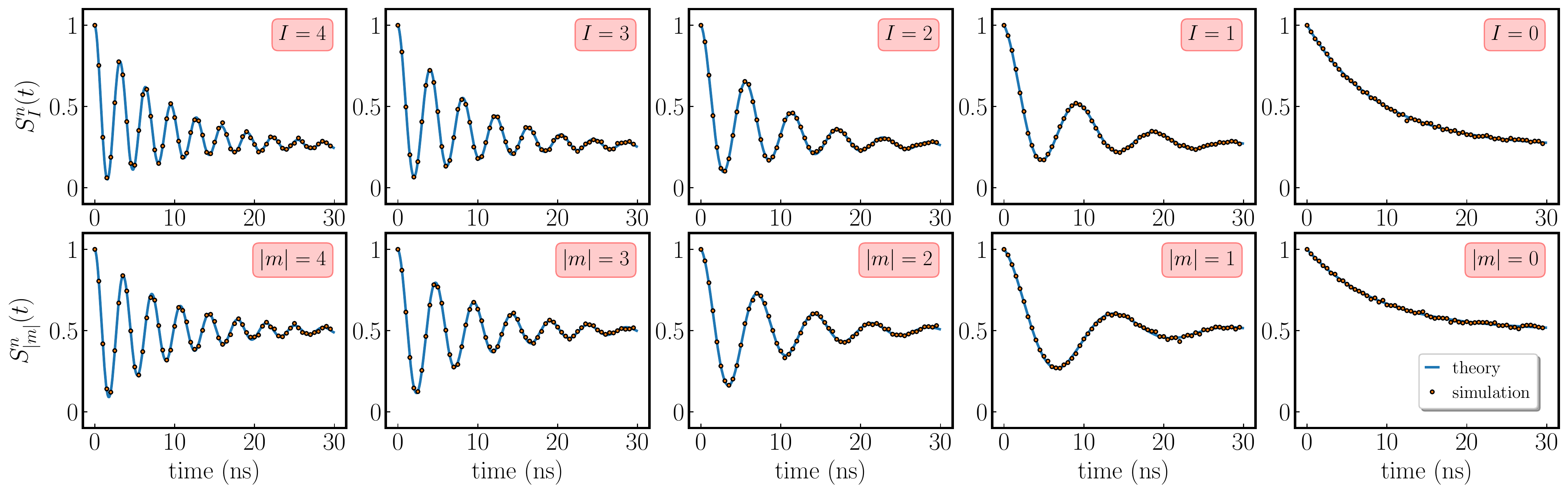}
\caption{Singlet state probability time evolution of individual pure nuclear initial states of 9,10-octalin$^+$ with $I = 4,3,2,1,0$ ($|m| = 4,3,2,1,0$ for all valid $I$), at zero (first row) and high (second row) magnetic field. The hyperfine coupling constant of 9,10-octalin was experimentally found to be $a=24.9G$. High magnetic field experiments were conducted at $B=0.3T$, and the g-factors of 9,10-octalin and PTP were approximated to be $g_1 = g_2 = 2.0028$. At zero magnetic field, thermal relaxation has $T_1 = T_2 = 9ns$ and at high field $T_1 = \infty$, $T_2 = 9ns$. Simulations were run on Qiskit \texttt{Aer} using the partitioned Hamiltonian simulation method (Appendix \ref{appendix:simplification}), and noise was simulated using the Kraus method.}
\label{fig:diff_octalin}
\end{figure}

\begin{figure}[H]
\centering
\includegraphics[width=0.95\textwidth]{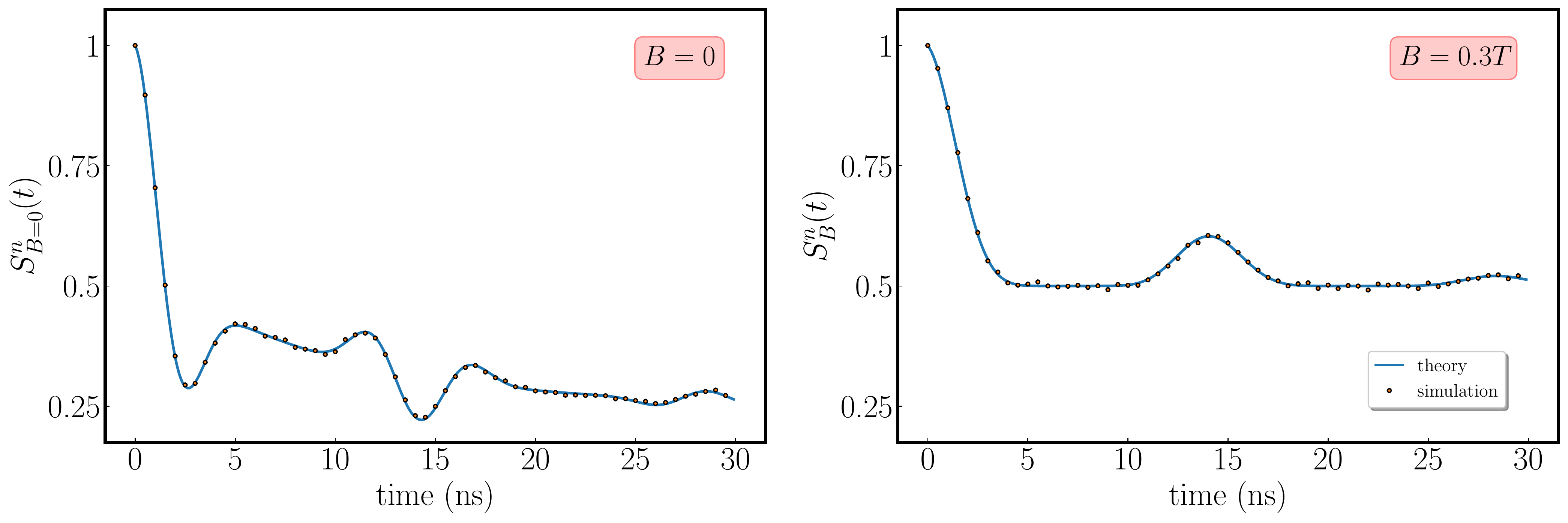}
\caption{Time dependence of the singlet state probabilities at zero (left) and high (right) magnetic field for the maximally mixed nuclear initial state of 9,10-octalin$^+$. Different pure nuclear initial states with $I = 4,3,2,1,0$ ($|m| = 4,3,2,1,0$) from Fig. \ref{fig:diff_octalin} were averaged using the weights from Table \ref{tab:statecounts}.}
\label{fig:zerohigh_octalin}
\end{figure}

The singlet state probability time evolution of the 9,10-octalin$^+$/PTP$^-$ radical pair simulated on Qiskit \texttt{Aer} using the partitioned Hamiltonian simulation method (Appendix \ref{appendix:simplification}) and the Kraus circuit for thermal relaxation is in excellent agreement with the analytic solution obtained by Bagryansky et. al \cite{bagryansky2000quantum} (Fig. \ref{fig:diff_octalin} and Fig. \ref{fig:zerohigh_octalin}). In general, in order to simulate the time evolution for 9,10-octalin$^+$ at an arbitrary magnetic field, simulation and averaging of 25 states is needed. However, it can be shown that the number of initial states can be reduced to just 5 under certain assumptions (Appendix \ref{appendix:octalin}). To that end, simulation results at zero and high magnetic fields for five pure nuclear initial state $\ket{I,m}$ for which there is a different probability pattern, are shown in Fig. \ref{fig:diff_octalin}. While at zero-field distinct probability curves are observed for nuclear initial states with different quantum numbers $I$, at high-field, it is the different magnetic numbers $|m|$ that result in distinct probability curves. Notably, oscillation frequencies depend on the HFC constant as well as the initial nuclear state.

Since chemistry experiments are performed at room temperature, which corresponds to effectively infinite temperature for a Hamiltonian system, the actual nuclear initial state is the maximally mixed state, where all states are equally likely. As such, to simulate the experimental time evolution, we classically build the weighted average of these different probability curves using the total nuclear state counts from Table \ref{tab:statecounts}. These probability curves at zero and high magnetic fields are depicted in Fig. \ref{fig:zerohigh_octalin} and show great agreement between theoretical and experimental results. 

\begin{figure}[H]
\centering
\includegraphics[width=10cm]{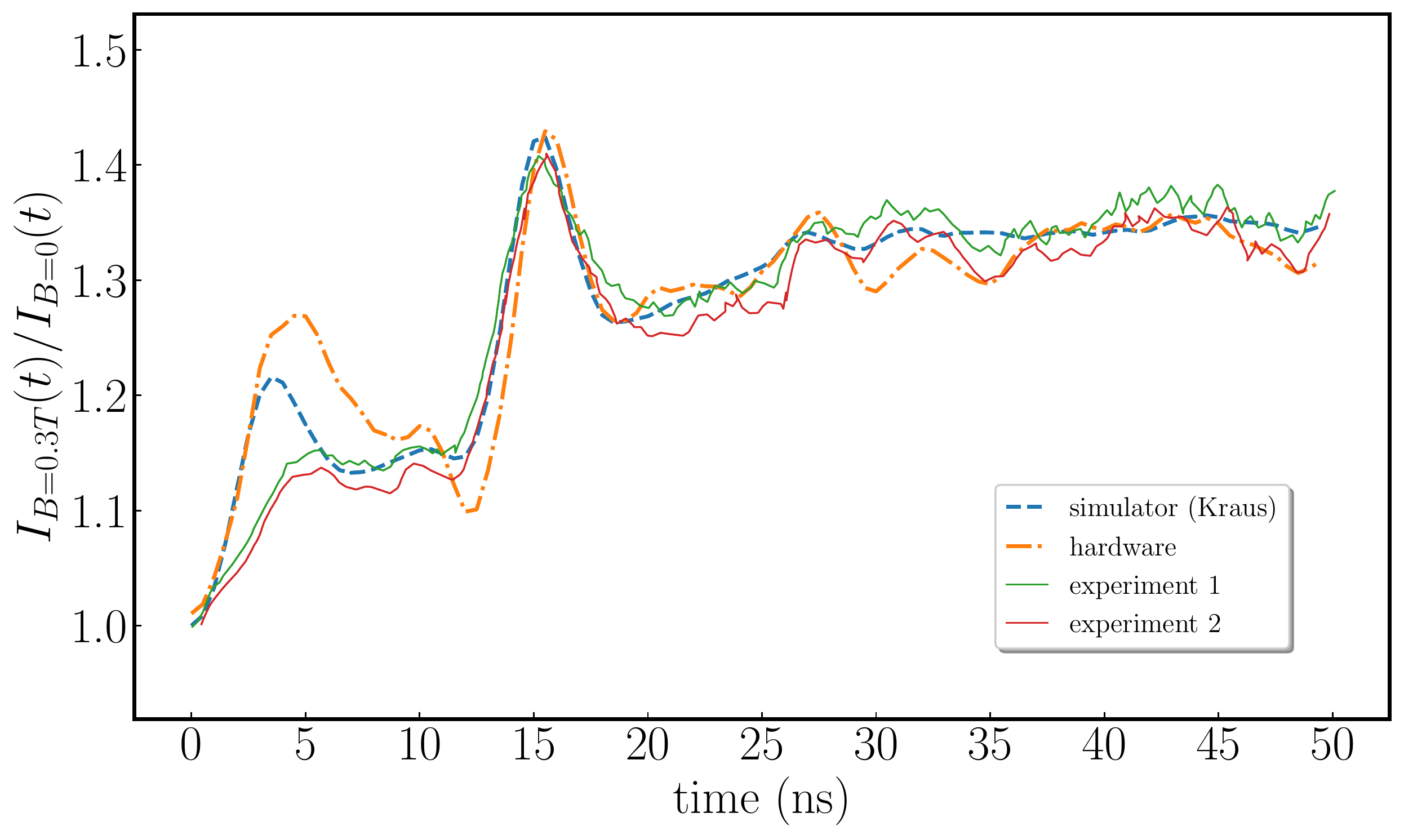}
\caption{Fluorescence intensity at high and zero magnetic field  ($I_B(t)/I_0(t)$ ratio) for 9, 10-octalin$^+$ as simulated with Kraus operators (in blue) vs. measured experimentally (in green and red) vs. simulated with quantum hardware (in orange). The experimental data was obtained from Bagryansky et al. \cite{bagryansky2000quantum}. Hardware experiments were run on the \textit{ibm\_lagos} quantum computer using the inherent qubit noise for thermal relaxation with full Hamiltonian simulation. Classical post-processing constants were $\theta=0.35$, $\tau_f=1.2ns$, $t_0=1ns$, $t_g=1ns$.}
\label{fig:endplot_octalin}
\end{figure}

Notably, the weighted averaging of the singlet state probability curves from Fig. \ref{fig:diff_octalin} is made possible by the specific basis selection during the Hamiltonian construction step. An alternative strategy allows us to initiate and simulate the Hamiltonian for the maximally mixed nuclear state (Appendix \ref{appendix:mixed}), but requires one ancilla qubit to be added to the circuit per nuclear qubit. This might be problematic for near-term quantum hardware where qubits are considered as an expensive resource. For systems with one group of magnetically equivalent nuclei such as 9,10-octalin, this concern can be circumvented by a careful basis selection. However, for systems with two or more groups of magnetically equivalent nuclei, the state counting arguments no longer apply, and the mixed-state initiation method needs to be employed as a subroutine to reduce the total number of simulated nuclear initial states. 
The singlet state probability evolution simulated on the quantum hardware with the inherent qubit noise match the experimental data better than simulator results, especially for longer times. The hardware and simulator results obtained for the $I_B(t)/I_0(t)$ ratio are compared with two sets of the experimental data obtained by Bagryansky et al. \cite{bagryansky2000quantum} (Fig. \ref{fig:endplot_octalin}). Experimental parameters $a, T_1, T_2$ and the classical post-processing parameter $\theta$ are determined by data fitting. Specifically, the HFC constant $a$ is obtained by fitting the time point at which the global maximum (second peak) occurs. This data-fit $a$ is found to be within a reasonable error range from our gas-phase quantum chemistry calculations. Interestingly, deviations between the hardware, simulator and experimental data can be observed at times less than $10ns$. The positions of the initial experimental and simulated peaks match reasonably well, but deviations in the peak intensity are observed. In the experiment, the first $10ns$ are not considered reliable, because the peaks observed during the initial $10ns$ are due to non-magnetically sensitive light emitted by other decaying excited states that are generated in the same radical pair solution upon initial irradiation. Radical pairs are typically separated by ca. $100nm$ in solution, and due to the relatively slow diffusion mechanism that facilitates radical recombination, exhibit longer lifetimes. Starting at ca. $8ns$, the main experimental peaks are very successfully simulated by the quantum hardware. The better match of hardware results to experimental data especially at later time points can be attributed to the fact that the simulator is considering the $T_1$ and $T_2$ relaxation only and can not account for other types of uncharacterized noise inherently present on the hardware.

\subsection{System with two groups of magnetically equivalent nuclei:  DMB\texorpdfstring{$^+$}{+}/PTP\texorpdfstring{$^-$}{-} radical pair}

\begin{figure}[H]
\centering
\includegraphics[width=\textwidth]{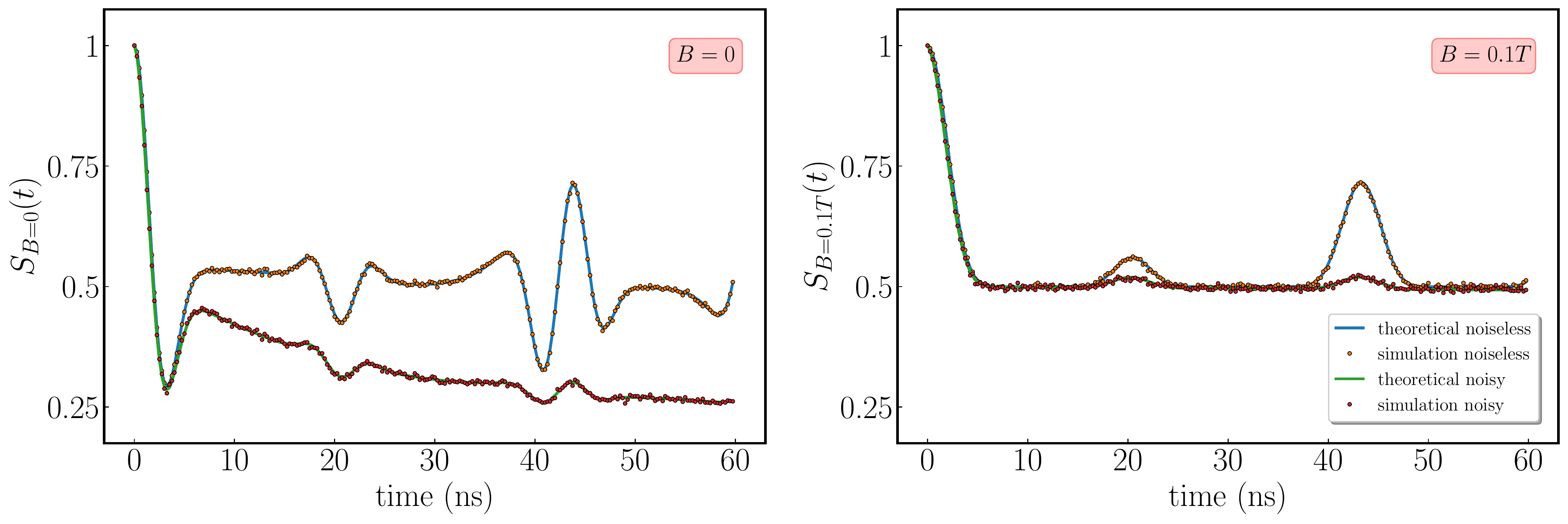}
\caption{Time dependence of the singlet state probabilities under Hamiltonian evolution with and without thermal relaxation at zero and high magnetic field for the maximally mixed nuclear initial state of DMB$^+$. Experimental constants are $a_1(2H)=6.5G$, $a_2(12H)=16.6G$; $T_1 = T_2 = 20ns$ at zero field, and $T_1 = 2000ns$, $T_2 = 20ns$ and $g_1=g_2=2.0028$ at high magnetic field of $B=0.1T$. Simulations were run using the partitioned mixed-state Hamiltonian simulation method with Qiskit \texttt{Aer}, and noise was simulated using the Kraus method.}
\label{fig:diff_dmb}
\end{figure}

The Hamiltonian simulations of the DMB$^+$/PTP$^-$ radical pair with the maximally mixed nuclear initial state were performed using, for the first time, the hyperfine coupling value as suggested by quantum chemistry calculations. The simulations show different singlet state probability patterns at zero and high magnetic fields, and under thermal noise (Fig. \ref{fig:diff_dmb}). At zero field, three sharp minima are observed at times ca. $2ns$, $20ns$ and $40ns$, respectively, followed by a peak at ca. $45ns$ and another small minimum at ca. $59 ns$. We pay close attention to the simulated peak and minimum pattern since it is determined by the HFC constants in our system. This pattern of peaks is a fingerprint of the system, which is why quantum beats have been developed into a spectroscopy technique \cite{ bagryansky2007review, molin2004spin_oscillations, molin1999quantum_beats, bagryansky2000quantum}. 

In the presence of paramagnetic relaxation introduced via the Kraus method, damping of oscillations is observed (Fig. \ref{fig:diff_dmb}). The singlet state probability decays as a function of time with a more rapid decay observed at zero magnetic field compared to high field. This is because experimentally, at zero field, $T_1$ and $T_2$ decay rates are nonzero, but at high field $T_1$ decay is negligible compared to $T_2$ decay. Additionally, the relaxation did not shift the peaks and minima found in the noiseless simulation, but only reduced their amplitudes. Notably, at high field,  convergence is reached when the singlet state population is 0.5, while at zero field, with all four spin states equally possible, the convergence is found at 0.25.

The hardware simulations for the DMB$^+$/PTP$^-$ radical pair were performed differently from the 9,10-octalin$^+$/PTP$^-$ radical pair. This is because the DMB$^+$/PTP$^-$ radical pair, having many more magnetically active nuclei, is not feasible to be completely treated by Hamiltonian simulation on the current quantum hardware (Fig. \ref{fig:hardware_encode} in Appendix \ref{appendix:dmb}). Instead, we obtained the coherent time evolution $S(t)$ of the singlet state probability from the simulator and mapped the decay from the decoherence of the qubits over time. 

\begin{figure}[H]
\centering
\includegraphics[width=10cm]{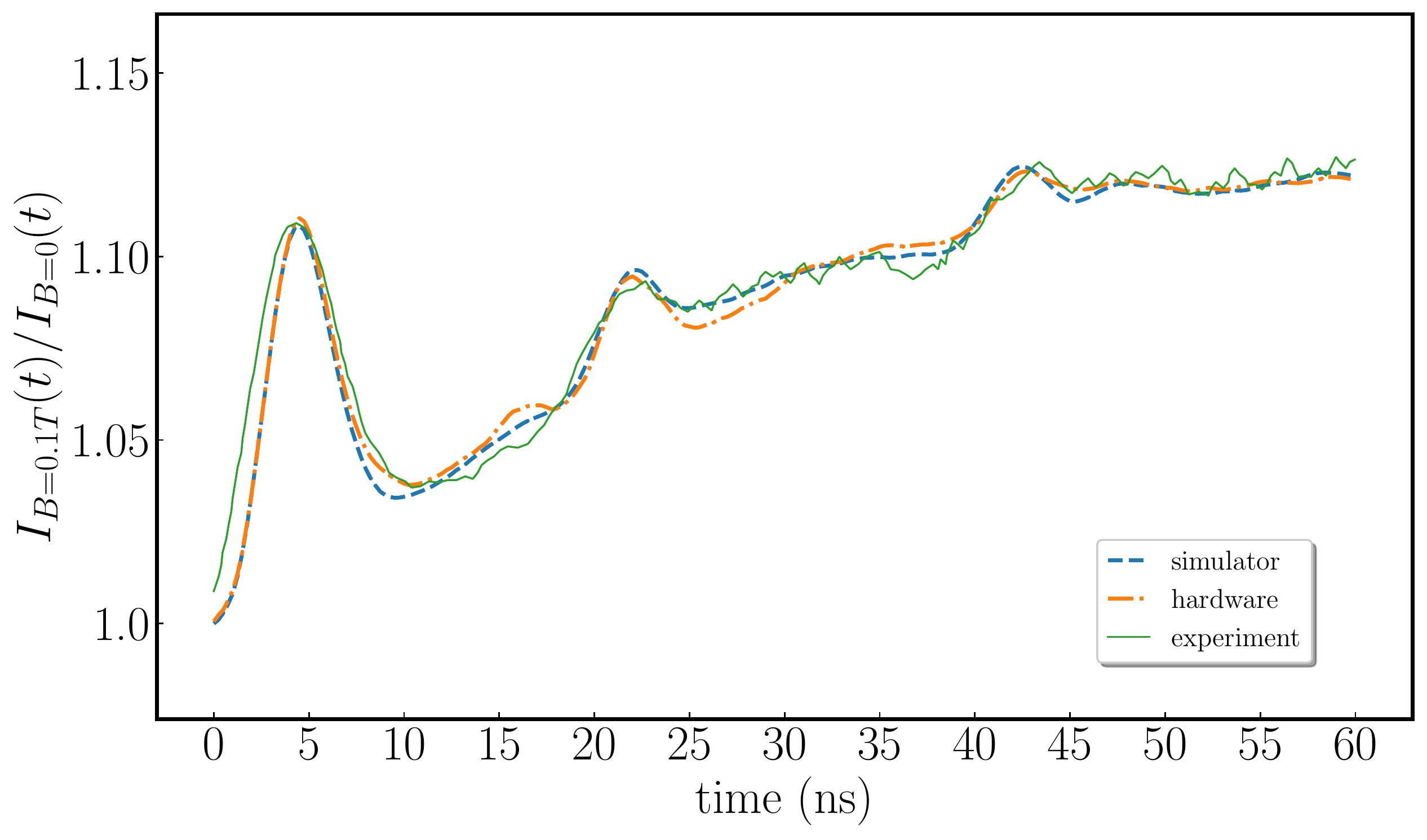}
\caption{Fluorescence intensity decay ($I_B(t)/I_0(t)$ ratio) for DMB$^+$ as simulated with a quantum simulator (in blue) and \textit{ibm\_lagos} quantum computer (in orange) vs. measured experimentally (in green). Experimental data was obtained by Bagryansky at al. \cite{bagryansky2005spin} and digitized \cite{Rohatgi2020}. Simulator results were obtained from the data depicted in Fig. \ref{fig:diff_dmb}. Classical post-processing constants were $\theta=0.132$, $\tau_f=1.2ns$, $t_0=1ns$, $t_g=1ns$.}
\label{fig:dmb_ratio}
\end{figure}

To compare our quantum simulator and hardware results with the experimental data obtained by Bagryansky et al. \cite{bagryansky2005spin}, we computed the ratio of fluorescence decay curves at high and zero magnetic fields  $I_B(t)/I_0(t)$. As shown in Fig. \ref{fig:dmb_ratio}, our quantum hardware and simulator curves are in very good agreement with experimental data, with respect to both the intensity and location of the peaks. The excellent agreement for times less than $10ns$ is due to a different solvent system used in the DMB$^+$/PTP$^-$ radical pair experiment from that in 9,10-octalin$^+$/PTP$^-$, which does not cause the excess fluorescence due to energy transfer from solvent to solute.

\section{Conclusions}

In this work, we demonstrate that quantum computers are naturally excellent platforms for the simulation of coherent time evolution and thermal relaxation effects in radical pairs undergoing the quantum-beat phenomena. The Hamiltonian simulations of the coherent time evolution of the radical pairs with exact parametrization of hyperfine coupling constants, different $g$-factors, and an arbitrary magnetic field strength, result in an exact analytic solution for systems with one or two groups of magnetically equivalent nuclei. We compare three different thermal relaxation simulation methods, i.e. Kraus channels, customizable Qiskit \texttt{Aer} noise models and the inherent qubit noise on quantum hardware, and find that the inherent qubit noise is able to reproduce the experimental results most accurately. Our results suggest that many sources of noise contribute to thermal relaxation of radical pairs and, surprisingly, this type of noise is inherently found in qubits, but not as much in noise models obtained with quantum simulators using only $T_1$ and $T_2$ noise (or any other of the currently built-in noise sources) or classical simulation methods. These results open up new ways to explore more complex spin-chemistry systems with nontrivial interactions and demonstrate that the inherent qubit noise can be an advantage over a classical computer for simulating open quantum systems in chemistry. 

\section{Acknowledgements}

All authors thank Dr. Gavin O Jones (IBM Quantum, IBM Research-Almaden) for useful chemistry discussions and comments. B.A.J. acknowledges the Aspen Center for Physics. M.T. thanks Prof. Tsachy Weissman (Stanford University) for helpful discussions on Hamiltonian analysis.

\bibliography{main}

\appendix
\section{Hyperfine coupling calculations for the 9,10-octalin and 2,3-dimethylbutane radical cations}
\label{appendix:hfc}

In this section, we provide additional data on the hyperfine coupling constant calculations for the two radical cations studied. 

\begin{table}[H]
    \begin{subtable}[h]{0.45\textwidth}
        \centering
        \begin{tabular}{ |c|c|c| } 
        \hline
        Atom ID & ($A_{iso}$/$\hbar$) (MHz)& ($A_{iso}$/$\hbar$) (mT) \\ 
        \hline\hline
        4 & 37.9699 & 1.617 \\ 
        5 & 110.874 & 4.7210 \\ 
        7 & 0.2224 & 0.0095  \\ 
        8 & 0.2173 & 0.0093  \\ 
        10 & 0.2194 & 0.0093 \\
        11 & 0.2208 & 0.0094  \\
        13 & 37.9519 & 1.616  \\
        14 & 110.887 & 4.7216  \\
        16 & 110.886 & 4.7215  \\
        17 & 37.9483 & 1.616  \\
        19 & 0.2189 & 0.0093  \\
        20 & 0.2213 & 0.0094  \\
        22 & 0.2219 & 0.0094  \\
        23 & 0.2178 & 0.0093  \\
        25 & 110.874 & 4.7210  \\
        26 & 37.9735 & 1.6169  \\
        \hline
        \end{tabular}
        \caption{9,10-octalin$^+$}
        \label{tab:orca_octalin}
    \end{subtable}
    \hfill
    \begin{subtable}[h]{0.45\textwidth}
        \centering
        \begin{tabular}{ |c|c|c| } 
        \hline
        Atom ID & ($A_{iso}$/$\hbar$) (MHz) & ($A_{iso}$/$\hbar$) (mT)\\ 
        \hline\hline
        2 & 2.9621 & 0.1261 \\ 
        3 & 114.3359 & 4.8684 \\
        4 & 29.0369 & 1.2364\\
        6 & 16.7892 & 0.7149 \\
        7 & 116.2778 & 4.9511 \\
        8 & 11.1075 & 0.4730 \\
        10 & 2.9763 & 0.1267 \\
        11 & 114.325 & 4.8680 \\
        12 & 28.9945 & 1.2346 \\
        14 & 11.1384 & 0.4743 \\
        15 & 16.7314 & 0.7124 \\
        16 & 116.3005 & 4.9521 \\
        18 & 22.9339 & 0.9764 \\
        20 & 22.9435 & 0.9769 \\
        \hline
        \end{tabular}
        \caption{2,3-dimethylbutane$^+$ (DMB)}
        \label{tab:orca_dmb}
     \end{subtable}
     \caption{Hyperfine coupling constants of the hydrogen atoms in the (a) 9,10-octalin$^+$ and (b) 2,3-dimethylbutane$^+$ radical cations as calculated using ORCA \cite{neese2020orca}. Labeled hydrogen atoms are shown in Fig. \ref{fig:atoms}.}
     \label{tab:orca}
\end{table}

\begin{figure}[H]
     \centering
     \begin{subfigure}[b]{0.45\textwidth}
         \centering
         \includegraphics[width=0.45\textwidth]{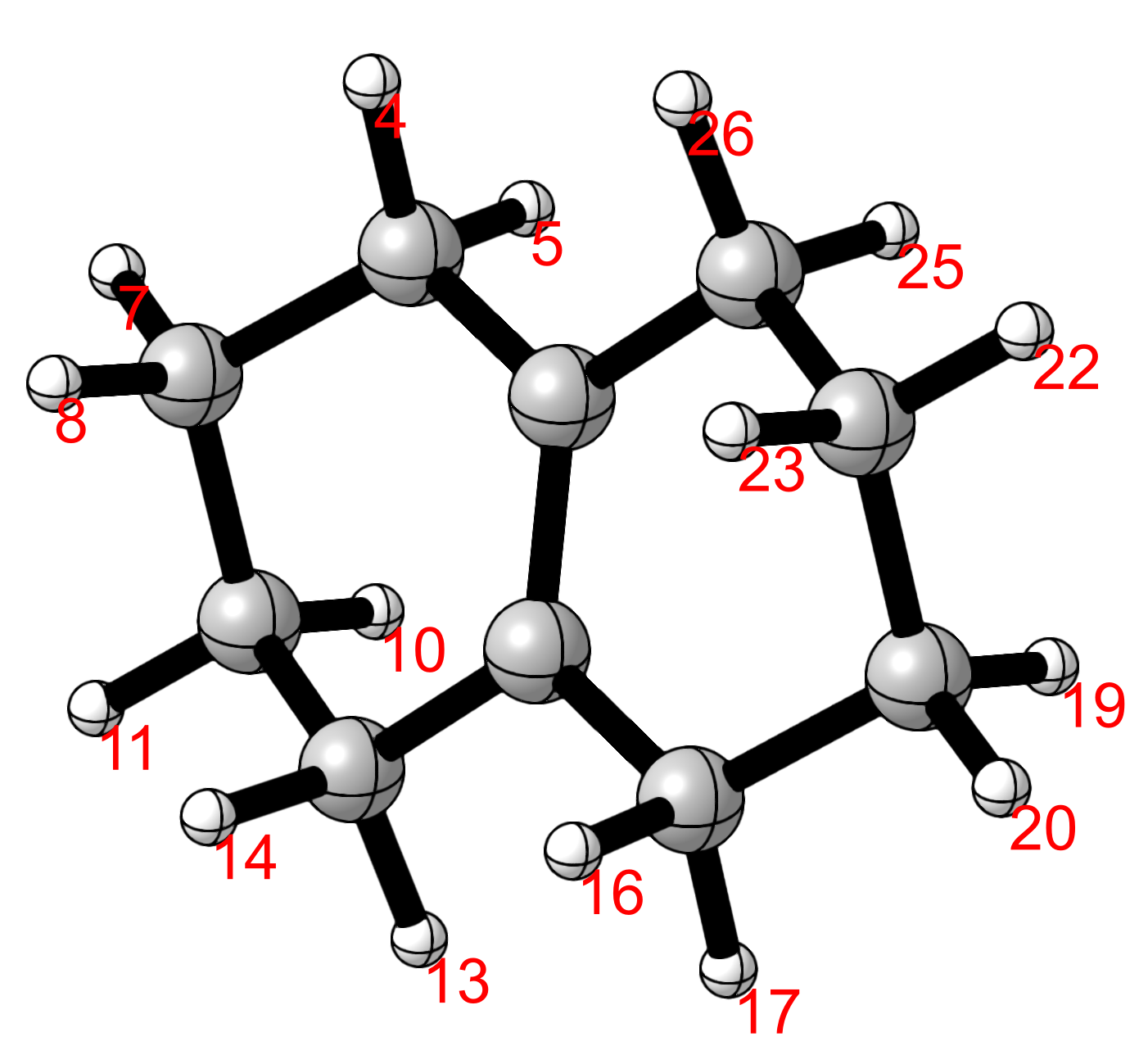}
         \caption{9,10-octalin$^+$}
         \label{fig:atoms_octalin}
     \end{subfigure}
     \hfill
     \begin{subfigure}[b]{0.49\textwidth}
         \centering
         \includegraphics[width=0.49\textwidth]{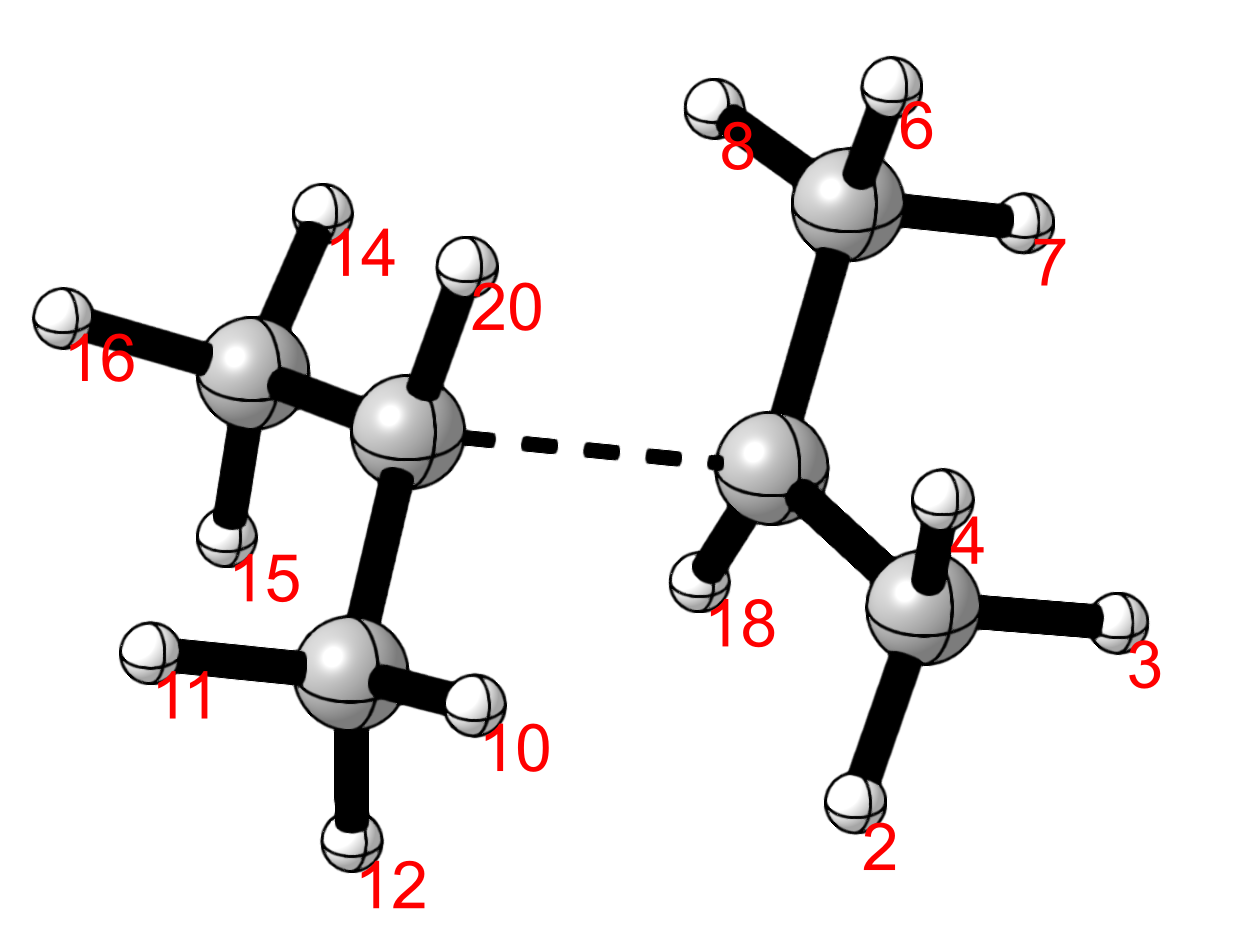}
         \caption{2,3-dimethylbutane$^+$}
         \label{fig:atoms_dmb}
     \end{subfigure}
        \caption{Optimized structure of the (a) 9,10-octalin and the (b) 2,3-dimethylbutane radical cations with labeled hydrogen atoms.}
        \label{fig:atoms}
\end{figure}

\section{Hamiltonian simulation of the 9,10-octalin\texorpdfstring{$^+$}{+}/PTP\texorpdfstring{$^-$}{-} radical pair}
\label{appendix:octalin}

\subsection{Full Hamiltonian construction}
\label{appendix:full_hamiltonian_octalin}
The full Hamiltonian matrix for 9,10-octalin$^+$/PTP$^-$ radical pair system represents 2 unpaired electrons and 8 proton hyperfine couplings. When we map this Hamiltonian onto a quantum computer, we  reserve one qubit per spin-1/2 particle in the system, i.e. 10 qubits, including all degeneracies. 

The overall Hamiltonian consists of three interaction terms: hyperfine coupling and the external magnetic field on the electronic subsystems of the radical cation ($B_1$) and anion ($B_2$), respectively, $H = H_{hfc} + H_{B_1} + H_{B_2}$. The external magnetic field terms $H_{B_1}$ and $H_{B_2}$ act independently on the electronic subsystems (cation and anion) in the radical pair, and, therefore, are diagonal in a basis obtained when the electronic subsystems are multiplied as a tensor product of two individual spin states. On the other hand, the total-spin states of 9,10-octalin$^+$, where 8 nuclear and 1 electronic spins are added together, are the eigenstates of $H_{hfc}$. Therefore the total Hamiltonian in either of the two bases will have off-diagonal elements. When we map to the quantum computer we use the first basis set. 

To compute the eigenenergies and eigenstates we count the number of total-spin states using the quantum mechanics rules of spin addition as shown in Table \ref{tab:addingspins}. There are $2^9=512$ states representing the degrees of freedom in the 9,10-octalin radical cation. 

\begingroup
\setlength{\tabcolsep}{6pt} 
\renewcommand{\arraystretch}{1} 
\begin{table}[H]
   \caption{Spin addition to count the total-spin states of 9,10-octalin. To progressively count the number of states with a given total spin, we start with a single spin-1/2 particle. Since adding another spin-1/2 both increases and decreases the spin by 1/2, we iteratively add all 8 nuclear spin-1/2s, and finally the remaining electronic subsystem in the last row.} 
   \label{tab:addingspins}
   \small
   \centering
   \begin{tabular}{l|c|c|c|c|c|c|c|c|c|r}
   \toprule
   \backslashbox{\textbf{Number}}{\textbf{Spin}}
    & \textbf{0} & \textbf{$\mathbf{\frac{1}{2}}$} & \textbf{1} & \textbf{$\mathbf{\frac{3}{2}}$} & \textbf{2} & \textbf{$\mathbf{\frac{5}{2}}$} & \textbf{3} & \textbf{$\mathbf{\frac{7}{2}}$} & \textbf{4} & \textbf{$\mathbf{\frac{9}{2}}$}\\ 
   \midrule
   1 & & 1 & & & & & & & &\\
   \midrule
   2 & 1 & & 1 & & & & & & &\\
   \midrule
   3 & & 2 & & 1 & & & & & &\\
   \midrule
   4 & 2 & & 3 & & 1 & & & & &\\
   \midrule
   5 & & 5 & & 4 & & 1 & & & &\\
   \midrule
   6 & 5 & & 9 & & 5 & & 1 & & &\\
   \midrule
   7 & & 14 & & 14 & & 6 & & 1 & &\\
   \midrule
   8 & 14 & & 28 & & 20 & & 7 & & 1 &\\
   \midrule
   9 & & 42 & & 48 & & 27 & & 8 & & 1\\
   \bottomrule
   \end{tabular}
\end{table}
\endgroup

For the first term in $H_{hfc}$ (equation \ref{eq:hfc}) out of 512 states and using row 9 of Table \ref{tab:addingspins}, we have: \\

\begin{center}
$42 \cdot 2 = 84$ states with $\bra{1/2}(\textbf{I}_{total}+\textbf{S}_1)^2\ket{1/2} = 1/2 \cdot 3/2 = 3/4$ \\
$48 \cdot 4 = 192$ states with $\bra{3/2}(\textbf{I}_{total}+\textbf{S}_1)^2\ket{3/2} = 3/2 \cdot 5/2 = 15/4$ \\
$27 \cdot 6 = 162$ states with $\bra{5/2}(\textbf{I}_{total}+\textbf{S}_1)^2\ket{5/2} = 5/2 \cdot 7/2 = 35/4$ \\
$8 \cdot 8 = 64$ states with $\bra{7/2}(\textbf{I}_{total}+\textbf{S}_1)^2\ket{7/2} = 7/2 \cdot 9/2 = 63/4$ \\
$1 \cdot 10 = 10$ states with $\bra{9/2}(\textbf{I}_{total}+\textbf{S}_1)^2\ket{9/2} = 9/2 \cdot 11/2 = 99/4$. \\ 
\end{center}
Similarly, for the second term in $H_{hfc}$ (equation \ref{eq:hfc}), we refer to row 8 of Table \ref{tab:addingspins}. Note that the multiplicity of 2 is due to the appended electronic spin-1/2. 
\begin{center}
$14 \cdot 1 \cdot 2= 28$ states with $\bra{0}\textbf{I}_{total}^2\ket{0} = 0 \cdot 1 = 0$ \\
$28 \cdot 3 \cdot 2= 168$ states with $\bra{1}\textbf{I}_{total}^2\ket{1} = 1 \cdot 2 = 2$ \\
$20 \cdot 5 \cdot 2= 200$ states with $\bra{2}\textbf{I}_{total}^2\ket{2} = 2 \cdot 3 = 6$ \\
$7 \cdot 7 \cdot 2= 98$ states with $\bra{3}\textbf{I}_{total}^2\ket{3} = 3 \cdot 4 = 12$ \\
$1 \cdot 9 \cdot 2= 18$ states with $\bra{4}\textbf{I}_{total}^2\ket{4} = 4 \cdot 5 = 20$. \\ 
\end{center}
The third and last term in $H_{hfc}$ (equation \ref{eq:hfc}) is $\textbf{S}_1^2$, which trivially acts on the electronic spin-1/2 and is $1/2 \cdot 3/2 = 3/4$. As a constant, it acts in the same way on every row of the Table \ref{tab:addingspins}. For the following, we use the notation $\ket{(I_{total} + S_1): I_{total}}$ to denote a quantum state of 9,10-octalin with total spin $I_{total}+S_1$ including the nuclear and electronic subsystems, and whose ``parent'' total nuclear spin is $I_{total}$. We obtain 9 different eigenenergies of $H_{hfc}$ as shown in Table \ref{tab:energies}. These would be the 9 eigenenergies of the total Hamiltonian with $B=0$.

\begin{table}[H]
   \caption{Component eigenenergies of $H_{hfc}$ in the total spin eigenbasis that includes 9 spin-1/2s of the 9,10-octalin radical cation.} 
   \label{tab:energies}
   \small
   \centering
   \begin{tabular}{l|c|c|c|c|c|c|c|c|r}
   \toprule
    $\ket{(I_{total} + S_1): I_{total}}$& $\ket{\frac{1}{2}: 0}$ & $\ket{\frac{1}{2}: 1}$ & $\ket{\frac{3}{2}: 1}$ & $\ket{\frac{3}{2}: 2}$& $\ket{\frac{5}{2}: 2}$& $\ket{\frac{5}{2}: 3}$& $\ket{\frac{7}{2}: 3}$& $\ket{\frac{7}{2}: 4}$& $\ket{\frac{9}{2}: 4}$\\
   \midrule
   counts & 28 & 56 & 112 & 80 & 120 & 42 & 56 & 8 & 10 \\
   \midrule \midrule
   $(\textbf{I}_{total} + \textbf{S}_1)^2$ & $\frac{3}{4}$ & $\frac{3}{4}$ & $\frac{15}{4}$ & $\frac{15}{4}$ & $\frac{35}{4}$ & $\frac{35}{4}$ & $\frac{63}{4}$ & 
   $\frac{63}{4}$ & $\frac{99}{4}$ \\
   \midrule
   $\textbf{I}_{total}^2$ & 0 & 2 & 2 & 6 & 6 & 12 & 12 & 
   20 & 20 \\
   \midrule
   $\textbf{S}_1^2$ & $\frac{3}{4}$ & $\frac{3}{4}$ & $\frac{3}{4}$ & $\frac{3}{4}$ & $\frac{3}{4}$ & $\frac{3}{4}$ & $\frac{3}{4}$ & $\frac{3}{4}$ & $\frac{3}{4}$ \\
   \midrule \midrule
   $H_{hfc} / a$ & 0 & -1 & 0.5 & -1.5 & 1 & -2 & 1.5 & -2.5 & 2 \\
   \bottomrule
   \end{tabular}
\end{table}

Since the initial state $\rho_0$ (equation \ref{eq:initial}) of the system is not an eigenstate, we use the theory of Clebsch-Gordan coefficients to re-express the electronic spins, entangled in the singlet state, in the eigenstate basis. On the quantum computer we use a unitary transformation to achieve this. Every pure nuclear initial state becomes coupled with the initial singlet state of the electronic subsystem. By identifying and enumerating the number of degenerate pure initial states, we run the Hamiltonian simulation on a very small subset of the basis states. We consistently order them as in Table \ref{tab:ordering}. The magnetic interaction terms of the Hamiltonian $H_{B_1}$ and $H_{B_2}$ are already diagonal in this basis.

The resulting Hamiltonian matrix can be simulated on a quantum computer without any simplifications. Let $\mathbb{1}_n$ denote an identity matrix of size $n \times n$, $\textrm{CG}_i$ the Clebsch-Gordan table in block matrix format for spin addition $i \oplus 1/2$, and $b_i = \frac{\mu_B}{2\hbar}g_iB$. $\Lambda_i$ are the total-spin basis eigenvalues arranged in the diagonal. Values under the brackets refer to the exact degeneracies. The resulting Hamiltonian appears in equations \ref{eq:b1} and \ref{eq:b2}.

\begin{align}
\label{eq:b1}
    \Tilde{H}_{hfc} &= 
    \begin{bmatrix}
    \underbrace{\textrm{CG}_4}_{\times 1} &  &  &  & \\
     & \underbrace{\textrm{CG}_3}_{\times 7} &  &  & \\
     &  & \underbrace{\textrm{CG}_2}_{\times 20} &  & \\
     &  &  & \underbrace{\textrm{CG}_1}_{\times 28} & \\
     &  &  &  & \underbrace{\textrm{CG}_0}_{\times 14}\\
    \end{bmatrix} 
    \begin{bmatrix}
    \underbrace{\Lambda_4}_{\times 1} &  &  &  & \\
     & \underbrace{\Lambda_3}_{\times 7} &  &  & \\
     &  & \underbrace{\Lambda_2}_{\times 20} &  & \\
     &  &  & \underbrace{\Lambda_1}_{\times 28} & \\
     &  &  &  & \underbrace{\Lambda_0}_{\times 14}\\
    \end{bmatrix} 
    \begin{bmatrix}
    \underbrace{\textrm{CG}_4}_{\times 1} &  &  &  & \\
     & \underbrace{\textrm{CG}_3}_{\times 7} &  &  & \\
     &  & \underbrace{\textrm{CG}_2}_{\times 20} &  & \\
     &  &  & \underbrace{\textrm{CG}_1}_{\times 28} & \\
     &  &  &  & \underbrace{\textrm{CG}_0}_{\times 14}\\
    \end{bmatrix} 
\end{align}
\begin{align}
\label{eq:b2}
    H = a
    \mathbb{1}_2 \otimes \Tilde{H}_{hfc} - b_1 \mathbb{1}_{512} \otimes Z - b_2 Z \otimes \mathbb{1}_{512}
\end{align}
Component eigenenergies $\Lambda_i$ of $H_{hfc}/a$ are arranged consistent with the state ordering in Table \ref{tab:ordering}.
\begin{align}
    \Lambda_4 &= \textrm{diag}\{2, 2, -2.5, 2, -2.5, 2, -2.5, 2, -2.5, 2, -2.5, 2, -2.5, 2, -2.5, 2, -2.5, 2\} \\
    \Lambda_3 &= \textrm{diag}\{1.5, 1.5, -2, 1.5, -2, 1.5, -2, 1.5, -2, 1.5, -2, 1.5, -2, 1.5\} \\
    \Lambda_2 &= \textrm{diag}\{1, 1, -1.5, 1, -1.5, 1, -1.5, 1, -1.5, 1\} \\
    \Lambda_1 &= \textrm{diag}\{0.5, 0.5, -1, 0.5, -1, 0.5\} \\
    \Lambda_0 &= \textrm{diag}\{0, 0\}
\end{align}

\begingroup
\setlength{\tabcolsep}{10pt} 
\renewcommand{\arraystretch}{1} 
\begin{table}[H]
   \caption{Basis state ordering for quantum computer simulations. Degeneracies are omitted for simplicity, and the exact state counts can be found in Table \ref{tab:addingspins}. Nuclear states are written as total spin states $\ket{I,m}$.} 
   \label{tab:ordering}
   \small
   \centering
   \begin{tabular}{ |p{3cm}|p{3.8cm}|p{3.8cm}|  }
    \hline
    \textbf{PTP electronic} & \textbf{9,10-Octalin nuclear} & \textbf{9,10-Octalin electronic} \\
    \hline
    \hline
    $\ket{\uparrow}$ & $\ket{4,4}$ & $\ket{\uparrow}$\\
    & & $\ket{\downarrow}$ \\
    & $\ket{4,3}$ & $\ket{\uparrow}$\\
    & & $\ket{\downarrow}$ \\
    & $\ket{4,2}$ & $\ket{\uparrow}$\\
    & & $\ket{\downarrow}$ \\
    & $\vdots$ & \\
    & $\ket{4,-4}$ & $\ket{\uparrow}$ \\ 
    & & $\ket{\downarrow}$ \\
    \cline{2-3}
    & $\ket{3,3}$ & $\ket{\uparrow}$\\
    & & $\ket{\downarrow}$ \\
    & $\vdots$ & \\
    & $\ket{3,-3}$ & $\ket{\uparrow}$ \\ 
    & & $\ket{\downarrow}$ \\
    \cline{2-3}
    & $\ket{2,2}$ & $\ket{\uparrow}$\\
    & & $\ket{\downarrow}$ \\
    & $\vdots$ & \\
    & $\ket{2,-2}$ & $\ket{\uparrow}$ \\ 
    & & $\ket{\downarrow}$ \\
    \cline{2-3}
     & $\ket{1,1}$ & $\ket{\uparrow}$\\
    & & $\ket{\downarrow}$ \\
    & $\ket{1,0}$ & $\ket{\uparrow}$ \\ 
    & & $\ket{\downarrow}$ \\
    & $\ket{1,-1}$ & $\ket{\uparrow}$ \\ 
    & & $\ket{\downarrow}$ \\
    \cline{2-3}
    & $\ket{0,0}$ & $\ket{\uparrow}$\\
    & & $\ket{\downarrow}$ \\
    \hline
    $\ket{\downarrow}$ & repeat & repeat\\
    & \vdots & \vdots \\
    \hline
    \end{tabular}
\end{table}
\endgroup

For 9,10-octalin$^+$, there are $25$ states that need to be averaged under arbitrary magnetic fields. It can be shown mathematically that for zero magnetic field, only states with different values of $I$ produce different $S(t)$ curves.  Similarly, for high magnetic fields under the equal g-factor assumption, only states with different magnetic numbers $|m|$ need to be considered. This reduces the number of initial states that need to be simulated from 25 to 5. Corresponding state counts are shown in Table \ref{tab:statecounts}. To initiate the quantum circuit in a selected nuclear state, we identify the state number in the basis ordering, convert it to binary, and use $X$ gates to flip the corresponding qubits into $\ket{1}$. Table \ref{tab:repstates} can be used as a reference to select particular initial states, provided that the basis states have been ordered according to Table \ref{tab:ordering}. The quantum circuit for the exemplary single instance of $\ket{I=4, m=2} \otimes \ket{S}$ from Table \ref{tab:repstates} is shown in Figure 13. 

\begin{table}[H]
   \caption{State counts obtained in accordance with row 8 of Table \ref{tab:addingspins}. Each cell indicates the number of total nuclear states $\ket{I,m}$ in the total-spin basis.} 
   \label{tab:statecounts}
   \small
   \centering
   \resizebox{0.49\textwidth}{!}{%
   \begin{tabular}{l|c|c|c|c|c|r}
   \toprule
    & $|m|=0$ & $|m|=1$ & $|m|=2$ & $|m|=3$ & $|m|=4$ & \textbf{zero field}\\ 
   \midrule
   $I=0$ & $1 \times 14$ & & & & & 14\\
   \midrule
   $I=1$ & $1 \times 28$ & $2 \times 28$ & & & & 84\\
   \midrule
   $I=2$ & $1 \times 20$ & $2 \times 20$ & $2 \times 20$ & & & 100\\
   \midrule
   $I=3$ & $1 \times 7$ & $2 \times 7$ & $2 \times 7$ & $2 \times 7$ & & 49\\
   \midrule
   $I=4$ & $1 \times 1$ & $2 \times 1$ & $2 \times 1$ & $2 \times 1$ & $2 \times 1$ & 9\\
   \midrule
   \textbf{high field} & 70 & 112 & 56 & 16 & 2 & \textbf{total: 256}\\
   \bottomrule
   \end{tabular}
   }%
\end{table}

\begingroup
\setlength{\tabcolsep}{6pt} 
\renewcommand{\arraystretch}{1.2} 
\begin{table}[H]
   \caption{Selecting representative initial states. For example, $\ket{I=4, m=2}$, is the second basis state starting the count from 0 in Table \ref{tab:ordering}, and can be initiated as $2=00000010 \equiv IIIIIIXI$ where the the qubit ordering uses the convention of $q_8q_7...q_1$.}
   \label{tab:repstates}
   \small
   \centering
   \resizebox{0.49\textwidth}{!}{%
   \begin{tabular}{l|c|c|c|c|c}
   \toprule
    $\ket{I,m}$& $\ket{4,4}$ & $\ket{4,3}$ & $\ket{4,2}$ & $\ket{4,1}$ & $\ket{4,0}$\\ 
    \midrule
    \textbf{decimal} & 0 & 1 & 2 & 3 & 4 \\
    \midrule
    \textbf{binary} & 00000000 &  00000001 & 00000010 & 00000011 & 00000100 \\
   \midrule \midrule
   $\ket{I,m}$& $\ket{4,-1}$ & $\ket{4,-2}$ & $\ket{4,-3}$ & $\ket{4,-4}$ & $\ket{3,3}$\\ 
   \midrule
    \textbf{decimal} & 5 & 6 & 7 & 8 & 9 \\
    \midrule
    \textbf{binary} & 00000101 &  00000110 & 00000111 & 00001000 & 00001001 \\
   \midrule \midrule
   $\ket{I,m}$& $\ket{3,2}$ & $\ket{3,1}$ & $\ket{3,0}$ & $\ket{3,-1}$ & $\ket{3,-2}$\\ 
   \midrule
    \textbf{decimal} & 10 & 11 & 12 & 13 & 14 \\
    \midrule
    \textbf{binary} & 00001010 &  00001011 & 00001100 & 00001101 & 00001110 \\
   \midrule \midrule
   $\ket{I,m}$& $\ket{3,-3}$ & $\ket{2,2}$ & $\ket{2,1}$ & $\ket{2,0}$ & $\ket{2,-1}$\\ 
   \midrule
    \textbf{decimal} & 15 & 58 & 59 & 60 & 61 \\
    \midrule
    \textbf{binary} & 00001111 &  00111010 & 00111011 & 00111100 & 00111101 \\
   \midrule \midrule
   $\ket{I,m}$& $\ket{2,-2}$ & $\ket{1,1}$ & $\ket{1,0}$ & $\ket{1,-1}$ & $\ket{0,0}$\\ 
   \midrule
    \textbf{decimal} & 62 & 158 & 159 & 160 & 242 \\
    \midrule
    \textbf{binary} & 00111110 &  10011110 & 10011111 & 10100000 & 11110010 \\
   \bottomrule
   \end{tabular}
    }%
\end{table}
\endgroup

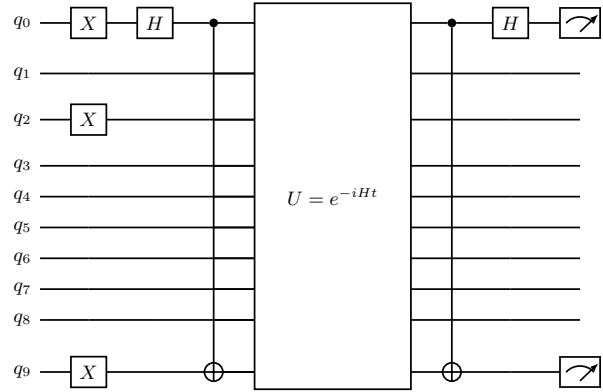
\begin{figure}[H]
    \centering
    \caption{Circuit for the Hamiltonian simulation of a specific nuclear case of $\ket{I=4, m=2}$. In this unsimplified method, one qubit is reserved per one spin-1/2 particle. $q_0$ is reserved for the electronic subsystem of 9,10-octalin and $q_9$ for PTP. $q_1$ through $q_8$ represent the nuclear subsystem.  Initial states can be selected by inserting $X$ gates according to the bit-string representation of the corresponding nuclear initial state.  Hamiltonian is provided in matrix format to Qiskit \texttt{HamiltonianGate} class for efficient gate decomposition.}
    \label{fig:circ1}
    \resizebox{0.45\textwidth}{!}{%
    \begin{quantikz}
    \lstick{$q_0$}& \gate{X} & \gate{H} & \ctrl{9} & \gate[10]{U = e^{-iHt}}[4cm] & \ctrl{9} & \gate{H} & \meter{} \\
    \lstick{$q_1$}& \qw & \qw & \qw & \qw & \qw & \qw & \qw\\
    \lstick{$q_2$}& \gate[]{X} & \qw & \qw & \qw & \qw & \qw & \qw \\
    \lstick{$q_3$}& \qw & \qw & \qw & \qw & \qw & \qw & \qw\\
    \lstick{$q_4$}& \qw & \qw & \qw & \qw & \qw & \qw & \qw\\
    \lstick{$q_5$}& \qw & \qw & \qw & \qw & \qw & \qw & \qw\\
    \lstick{$q_6$}& \qw & \qw & \qw & \qw & \qw & \qw & \qw\\
    \lstick{$q_7$}& \qw & \qw & \qw & \qw & \qw & \qw & \qw\\
    \lstick{$q_8$}& \qw & \qw & \qw & \qw & \qw & \qw & \qw\\
    \lstick{$q_9$}& \gate{X} & \qw & \targ{} & \qw & \targ{} & \qw  & \meter{}
    \end{quantikz}
    }%
\end{figure}

\subsection{Mixed state initiation}
\label{appendix:mixed}
It is possible to run the time evolution on the exact maximally mixed initial nuclear state without having to run multiple simulations on different initial states. Although beneficial, this method does not allow for circuit simplification, and requires as many additional ancilla qubits. 

\begin{figure}[H]
    \centering
    \caption{Mixed state Hamiltonian simulation circuit. We maximally entangle the nuclear qubits with an ancilla system of the same size.}
    \label{fig:circ2}
    \resizebox{0.65\textwidth}{!}{%
    \begin{quantikz}
    \lstick{$q_0$}& \gate{X} & \gate{H} & \qw & \qw & \qw & \qw & \qw & \qw & \qw & \ctrl{9} & \gate[10]{U = e^{-iHt}} & \ctrl{9} & \gate{H} & \meter{} \\
    \lstick{$q_1$}& \qw & \targ{} & \qw & \qw & \qw & \qw & \qw & \qw & \qw & \qw & \qw & \qw & \qw & \qw\\
    \lstick{$q_2$}& \qw & \qw & \targ{} & \qw & \qw & \qw & \qw & \qw & \qw & \qw & \qw & \qw & \qw & \qw \\
    \lstick{$q_3$}& \qw & \qw & \qw & \targ{} & \qw & \qw & \qw & \qw & \qw & \qw & \qw & \qw & \qw & \qw\\
    \lstick{$q_4$}& \qw & \qw & \qw  & \qw & \targ{} & \qw & \qw & \qw & \qw & \qw & \qw & \qw & \qw & \qw\\
    \lstick{$q_5$}& \qw & \qw & \qw & \qw & \qw & \targ{} & \qw & \qw & \qw & \qw & \qw & \qw & \qw & \qw\\
    \lstick{$q_6$}& \qw & \qw & \qw & \qw & \qw & \qw & \targ{} & \qw & \qw & \qw & \qw & \qw & \qw & \qw\\
    \lstick{$q_7$}& \qw & \qw & \qw & \qw & \qw & \qw & \qw & \targ{} & \qw & \qw & \qw & \qw & \qw & \qw\\
    \lstick{$q_8$}& \qw & \qw & \qw & \qw & \qw & \qw & \qw & \qw & \targ{} & \qw & \qw & \qw & \qw & \qw\\
    \lstick{$q_9$}& \gate{X} & \qw& \qw & \qw & \qw & \qw & \qw & \qw & \qw & \targ{} & \qw & \targ{} & \qw  & \meter{} \\
    \lstick{$q_1^A$}& \gate{H} & \ctrl{-9} & \qw & \qw & \qw & \qw & \qw & \qw & \qw & \qw & \qw & \qw & \qw & \qw\\
    \lstick{$q_2^A$}& \gate{H} & \qw & \ctrl{-9} & \qw & \qw & \qw & \qw & \qw & \qw & \qw & \qw & \qw & \qw & \qw\\
    \lstick{$q_3^A$}& \gate{H} & \qw & \qw & \ctrl{-9} & \qw & \qw & \qw & \qw & \qw & \qw & \qw & \qw & \qw & \qw\\
    \lstick{$q_4^A$}& \gate{H} & \qw & \qw & \qw & \ctrl{-9} & \qw & \qw & \qw & \qw & \qw & \qw & \qw & \qw & \qw\\
    \lstick{$q_5^A$}& \gate{H} & \qw & \qw & \qw & \qw & \ctrl{-9} & \qw & \qw & \qw & \qw & \qw & \qw & \qw & \qw\\
    \lstick{$q_6^A$}& \gate{H} & \qw & \qw & \qw & \qw & \qw & \ctrl{-9} & \qw & \qw & \qw & \qw & \qw & \qw & \qw\\
    \lstick{$q_7^A$}& \gate{H} & \qw & \qw & \qw & \qw & \qw & \qw & \ctrl{-9} & \qw & \qw & \qw & \qw & \qw & \qw\\
    \lstick{$q_8^A$}& \gate{H} & \qw & \qw & \qw & \qw & \qw & \qw & \qw & \ctrl{-9} & \qw & \qw & \qw & \qw & \qw\\
    \end{quantikz}
    }%
\end{figure}
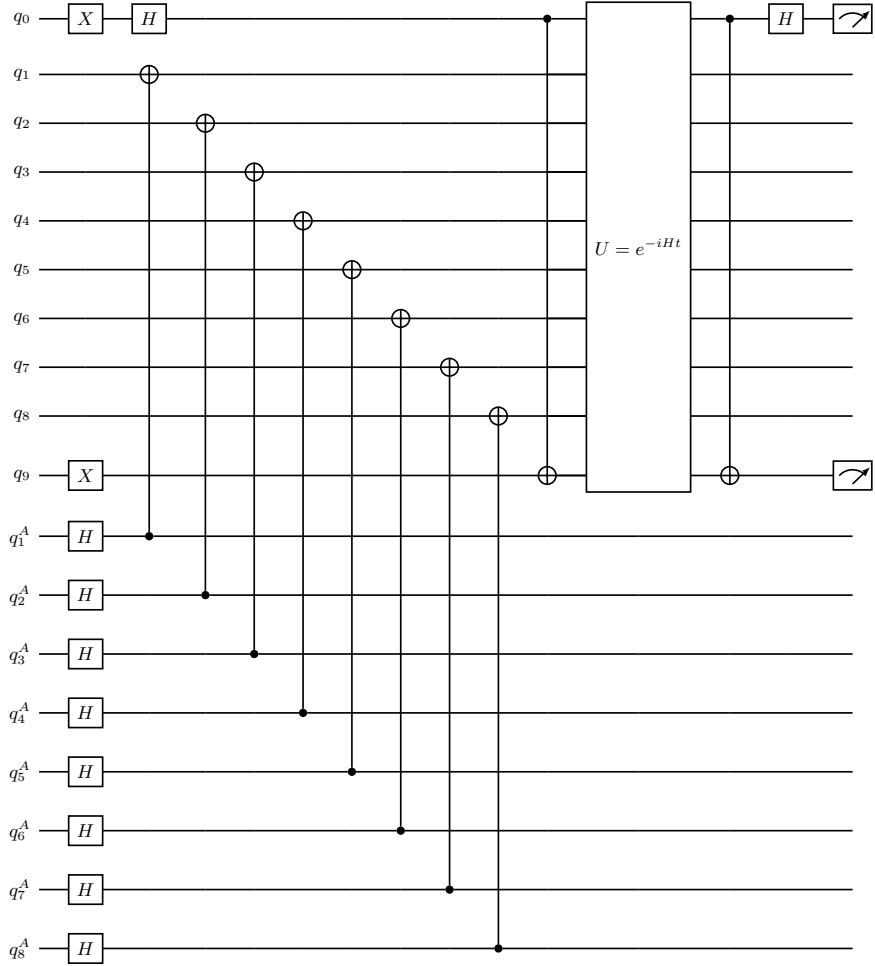

To initiate a maximally mixed (nuclear) state on a quantum circuit, we make use of quantum state purification techniques. Quantum state purification \cite{wilde2013quantum} states that any mixed state can be represented as a pure state if it is entangled with an appropriate ancilla system $A$. Specifically, the purification of the maximally mixed state is the maximally entangled state:
\begin{align}
    \ket{\Psi} = \frac{1}{\sqrt{2^N}} \sum_{n=0}^{2^N-1} \ket{n} \ket{n}_A,
\end{align}
which allows us to recover the maximally mixed state when we trace out the ancilla system $A$: 
\begin{align}
    \rho_0 = \frac{\mathbb{1}}{2^N} \otimes \ket{S} \bra{S} = \textrm{Tr}_A \bigg\{ \frac{1}{2^N} \sum_{n=0}^{2^N-1} \sum_{m=0}^{2^N-1} \ket{n} \ket{n}_A \bra{m} \bra{m}_A \bigg\} \otimes \ket{S} \bra{S}
\end{align}
The circuit that initiates $\ket{\Psi}$ for 9,10-octalin$^+$ with $N=8$ is shown in Fig. \ref{fig:circ2}. 

\subsection{Simplifications to the Hamiltonian simulation circuit}
\label{appendix:simplification}

For 9,10-octalin$^+$/PTP$^-$ radical pair Hamiltonian simulation, 5 qubits are enough (instead of 8) to represent the 25 nuclear states $\ket{I,m}$ of the 9,10-octalin$^+$/PTP$^-$ radical pair. However, to make sure our Hamiltonian matrix is of $2^n$ size ($2^5 = 32$), we append $7\cdot2$ zeros (since $32-25=7$) to achieve the right size. The desired initial state can be selected from Table \ref{tab:statecounts_reduced}.

\begin{align}
    \Tilde{H}_{hfc} &= 
    \begin{bmatrix}
    \textrm{CG}_4 &  &  &  &  & \\
     & \textrm{CG}_3 &  &  &  & \\
     &  & \textrm{CG}_2 &  &  & \\
     &  &  & \textrm{CG}_1 &  & \\
     &  &  &  & \textrm{CG}_0 & \\
     & & & & & \mathbb{0}_{14}
    \end{bmatrix} 
    \begin{bmatrix}
    \Lambda_4 &  &  &  & &\\
     & \Lambda_3 &  &  & &\\
     &  & \Lambda_2 &  & &\\
     &  &  & \Lambda_1 & &\\
     &  &  &  & \Lambda_0 &\\
     & & & & & \mathbb{0}_{14}
    \end{bmatrix} 
    \begin{bmatrix}
    \textrm{CG}_4 &  &  &  & &\\
     & \textrm{CG}_3 &  &  & &\\
     &  & \textrm{CG}_2 &  & &\\
     &  &  & \textrm{CG}_1 & &\\
     &  &  &  & \textrm{CG}_0 &\\
     & & & & & \mathbb{0}_{14}
    \end{bmatrix} 
\end{align}

\begin{align}
    H = a
    \mathbb{1}_2 \otimes \Tilde{H}_{hfc} - b_1 \mathbb{1}_{64} \otimes Z - b_2 Z \otimes \mathbb{1}_{64}
\end{align}

where $\mathbb{1}_n$ ($\mathbb{0}_n$) denotes an identity (zero) matrix of size $n \times n$, $\textrm{CG}_i$ the spin addition tables in block matrix format for adding $i \oplus 1/2$, $b_i = \frac{\mu_B}{2\hbar}g_iB$, and $\Lambda_i$ are the component eigenenergies arranged in the diagonal. \\

\begingroup
\setlength{\tabcolsep}{6pt} 
\renewcommand{\arraystretch}{1.2} 
\begin{table}[H]
   \caption{Selecting initial states for the Hamiltonian matrix that omits degeneracies.}
   \label{tab:statecounts_reduced}
   \small
   \centering
   \resizebox{0.43\textwidth}{!}{%
   \begin{tabular}{l|c|c|c|c|r}
   \toprule
    $\ket{I,m}$& $\ket{4,4}$ & $\ket{4,3}$ & $\ket{4,2}$ & $\ket{4,1}$ & $\ket{4,0}$\\ 
    \midrule
    \textbf{decimal} & 0 & 1 & 2 & 3 & 4 \\
    \midrule
    \textbf{binary} & 00000 &  00001 & 00010 & 00011 & 00100 \\
   \midrule \midrule
   $\ket{I,m}$& $\ket{4,-1}$ & $\ket{4,-2}$ & $\ket{4,-3}$ & $\ket{4,-4}$ & $\ket{3,3}$\\ 
   \midrule
    \textbf{decimal} & 5 & 6 & 7 & 8 & 9 \\
    \midrule
    \textbf{binary} & 00101 &  00110 & 00111 & 01000 & 01001 \\
   \midrule \midrule
   $\ket{I,m}$& $\ket{3,2}$ & $\ket{3,1}$ & $\ket{3,0}$ & $\ket{3,-1}$ & $\ket{3,-2}$\\ 
   \midrule
    \textbf{decimal} & 10 & 11 & 12 & 13 & 14 \\
    \midrule
    \textbf{binary} & 01010 &  01011 & 01100 & 01101 & 01110 \\
   \midrule \midrule
   $\ket{I,m}$& $\ket{3,-3}$ & $\ket{2,2}$ & $\ket{2,1}$ & $\ket{2,0}$ & $\ket{2,-1}$\\ 
   \midrule
    \textbf{decimal} & 15 & 16 & 17 & 18 & 19 \\
    \midrule
    \textbf{binary} & 01111 &  10000 & 10001 & 10010 & 10011 \\
   \midrule \midrule
   $\ket{I,m}$& $\ket{2,-2}$ & $\ket{1,1}$ & $\ket{1,0}$ & $\ket{1,-1}$ & $\ket{0,0}$\\ 
   \midrule
    \textbf{decimal} & 20 & 21 & 22 & 23 & 24 \\
    \midrule
    \textbf{binary} & 10100 &  10101 & 10110 & 10111 & 11000 \\
   \bottomrule
   \end{tabular}
   }%
\end{table}
\endgroup

There is another approach that results in the lowest possible number of qubits, but it requires additional circuits per initial state of interest to be constructed. This observation follows from the structure of Clebsch-Gordan matrices for spin addition $i \oplus 1/2$. Specifically, the Hamiltonian consists of diagonal blocks of size at most $2 \times 2$ and any pure initial state of form $\ket{I,m} \otimes \ket{S}$ interacts with at most 2 blocks. The two diagonal blocks that the initial state interacts with are chosen and separate circuits are built for each of those cases. Recalling that we only need one representative nuclear initial state for different values of $I$ ($|m|$) at zero (high) magnetic field, the simplest partitioning of the Hamiltonian is parameterized in a way where $x, y$ are Clebsch-Gordan coefficients, and $\lambda_1, \lambda_2$ are component eigenenergies. For simplicity, it is sufficient to parameterize the evolution of nuclear states of form $\ket{I, m=I}.$ Corresponding parameters are shown in Table \ref{tab:partitionedparams}.

\begin{align}
\label{eq:hfc_partitioned}
    H_{hfc} = a
    \begin{bmatrix}
1 & & \\
 & x & y \\
 & y & -x
\end{bmatrix}
\begin{bmatrix}
\lambda_1 & & \\
 & \lambda_1 &  \\
 &  & \lambda_2
\end{bmatrix}
\begin{bmatrix}
1 & & \\
 & x & y \\
 & y & -x
\end{bmatrix}
\end{align}
\begin{align}
    H = \mathbb{1}_2 \otimes \Bigg(
    \begin{bmatrix}
    H_{hfc} & \\
    & 0
    \end{bmatrix}
    + \mathbb{1}_2 \otimes \frac{b_1}{2}
    \begin{bmatrix}
    -1 & \\
    & 1
    \end{bmatrix} \Bigg)
    + \frac{b_2}{2}
    \begin{bmatrix}
    -1 & \\
    & 1 
    \end{bmatrix}
    \otimes \mathbb{1}_4
\end{align}

\begingroup
\setlength{\tabcolsep}{6pt} 
\renewcommand{\arraystretch}{1.2} 
\begin{table}[H]
   \caption[Parametrization of the partitioned Hamiltonian]{Partitioned Hamiltonian parametrization. Hamiltonian simulation given any nuclear initial state $\ket{I,m=I}$ of the 9,10-octalin radical cation, can be set up by inserting the corresponding Clebsch-Gordan coefficients $x,y$ and the eigenenergies $\lambda_1,\lambda_2$ into equation \ref{eq:hfc_partitioned}.} 
   \label{tab:partitionedparams}
   \small
   \centering
   \resizebox{0.42\textwidth}{!}{%
   \begin{tabular}{ p{1.2cm}|p{1.2cm}|p{1.2cm}|p{1.2cm}|p{1.2cm} }
   \toprule
    $\ket{I,m}$& $x$ & $y$ & $\lambda_1$ & $\lambda_2$ \\
   \midrule \midrule
   $\ket{4,4}$& $\sqrt{1/9}$ & $\sqrt{8/9}$ & $2$ & $-2.5$ \\
   \midrule
   $\ket{3,3}$& $\sqrt{1/7}$ & $\sqrt{6/7}$ & $1.5$ & $-2$ \\
   \midrule
   $\ket{2,2}$& $\sqrt{1/5}$ & $\sqrt{4/5}$ & $1$ & $-1.5$ \\
   \midrule
   $\ket{1,1}$& $\sqrt{1/3}$ & $\sqrt{2/3}$ & $0.5$ & $-1$ \\
   \midrule
   $\ket{0,0}$& - & - & $0$ & $0$ \\
   \bottomrule
   \end{tabular}
   }%
\end{table}
\endgroup

The Hamiltonian can also be expressed as a linear combination of Pauli matrices. For each Pauli tensor product, an explicit circuit is constructed, and Trotterization is performed on the non-commuting terms for the overall Hamiltonian simulation.
\begin{align}
H &= \bigg[ \frac{\lambda_1}{4} + \frac{\lambda_1 + \lambda_2}{4} \bigg] \; III
+ \bigg[ \frac{\lambda_1}{4} + \frac{(\lambda_1 - \lambda_2) (x^2 - y^2)}{4} \bigg] \; IZI
+ \bigg[ \frac{\lambda_1}{4} - \frac{(\lambda_1 - \lambda_2) (x^2 - y^2)}{4} - \frac{b_1}{2} \bigg] \; IIZ\\
&+ \bigg[ \frac{\lambda_1}{4} - \frac{\lambda_1 + \lambda_2}{4} \bigg] \; IZZ
+ \frac{(\lambda_1 - \lambda_2)xy}{2} \; IXX
+ \frac{(\lambda_1 - \lambda_2)xy}{2} \; IYY
-\frac{b_2}{2} ZII \nonumber
\end{align}

\begin{figure}[H]
    \centering
    \caption[KAK decomposition circuit for the partitioned Hamiltonian]{KAK decomposition circuit for the partitioned Hamiltonian simulation. $H_{hfc} + H_{B_1}$ acting on $q_0, q_1$ is decomposed into the $U3$ parameters using \texttt{two\_qubit\_cnot\_decompose} function in Qiskit. $H_{B_2}$ is a Z-rotation on $q_2$.}
    \label{fig:kak_partitioned}
    \resizebox{\textwidth}{!}{%
    \begin{quantikz}
    \lstick{$q_0$}& \gate{X} & \gate{H} & \ctrl{2} & \gate{U3(\theta_{a},\phi_{a},\lambda_{a})}\gategroup[wires=2,steps =7,style={dashed, rounded corners,fill=blue!20, inner xsep=2pt}, background]{$H_{hfc}+H_{B_1}$} & \ctrl{1} & \gate{U3(\theta_{c},\phi_{c},\lambda_{c})} & \ctrl{1} & \gate{U3(\theta_{e},\phi_{e},\lambda_{e})} & \ctrl{1} & \gate{U3(\theta_{g},\phi_{g},\lambda_{g})} & \ctrl{2} & \gate{H} & \meter{} \\
    \lstick{$q_1$}& \qw & \qw & \qw & \gate{U3(\theta_{b},\phi_{b},\lambda_{b})} & \targ{} & \gate{U3(\theta_{d},\phi_{d},\lambda_{d})} & \targ{} & \gate{U3(\theta_{f},\phi_{f},\lambda_{f})} & \targ{} & \gate{U3(\theta_{h},\phi_{h},\lambda_{h})} & \qw & \qw & \qw\\
    \lstick{$q_2$}& \gate{X} & \qw & \targ{} & \qw & \qw & \qw & \gate{R_Z(\theta)}\gategroup[wires=1,steps =1,style={dashed, rounded corners,fill=red!20, inner xsep=2pt, }, background, label style={label position=below,anchor= north,yshift=-0.2cm}]{$H_{B_2}$} & \qw & \qw & \qw  & \targ{} & \qw & \meter{}
    \end{quantikz}
    }%
\end{figure}
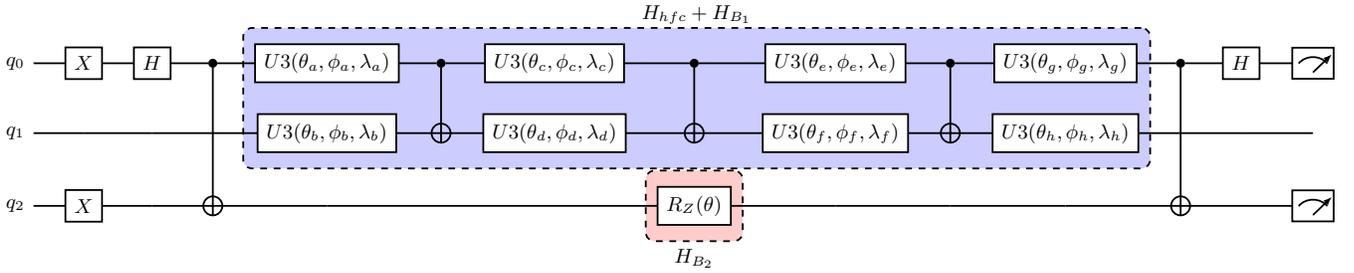

Additionally, the Hamiltonian can be decoupled into two terms, $H_{hfc}+H_{B_1}$ and $H_{B_2}$, that act on different qubits. The first term is acting on two qubits, and its corresponding unitary can be decomposed into three sets of CNOT gates and four sets of local operations using the KAK decomposition \cite{kraus2001optimal, vatan2004optimal, vidal2004universal}. The second term is implemented with a Z-rotation gate (Fig. \ref{fig:kak_partitioned}).

\section{Hamiltonian simulation of the 2,3-dimethylbutane\texorpdfstring{$^+$}{+}/PTP\texorpdfstring{$^-$}{-} radical pair}
\label{appendix:dmb}
The 2,3-dimethylbutane (DMB) radical cation has two groups of magnetically equivalent nuclei, 2 protons with the hyperfine coupling (HFC) constant $a_1$ and 12 protons with $a_2$. The Hamiltonian setup for radical cations with two groups of magnetically equivalent nuclei comes with a technical challenge - the construction of a different basis set for the Hamiltonian matrix compared to that of 9,10-octalin radical cation. Methods discussed here build upon Appendix \ref{appendix:octalin}. The simplest basis set in which the Hamiltonian can be mapped onto a quantum computer and its corresponding state ordering are listed in Table \ref{tab:ordering_dmb}. Notably, the component eigenbases of both $H_1$ and $H_2$ include the electronic spin-1/2 in their respective total-spin states making $H_{hfc}=H_1+H_2$ not diagonal in this basis. To circumvent that we obtain the component eigenenergies of $H_1$ by adding the electronic spin to nuclear spin $I_1$ first and then transform our bases using Clebsch-Gordan coefficients. We repeat this for $H_2$ with $I_2$ and obtain the basis and eigenenergies shown in Table \ref{tab:ordering_dmb}.

Given that the nuclear initial state is the maximally mixed state, we need run this Hamiltonian for every different $\ket{I_2, m_2} \ket{I_1, m_1}$ initial state combination. In this basis every nuclear initial state produces a different measurement outcome $S(t)$. For a radical cation with $N$ nuclei, this increases the required number of simulations from $O(N)$ to $O(N^4)$. Fortunately, because of the underlying symmetries of the full Hamiltonian, it block-diagonalizes into subsections corresponding to conserved quantum numbers. This enables us to partition the Hamiltonian matrix into blocks that we can simulate separately. To run Hamiltonian simulation on a quantum computer, we append all-zero sub-matrices to round up the size of the matrices to the nearest power of 2, if needed. We make use of the mixed state method for each block to avoid the polynomial increase in complexity. Eventually, the effect of the appended zeros can be removed, and the probabilities obtained from different simulations can be combined with the correct degeneracies using classical post-processing steps.

The Hamiltonian matrix construction for $I_2=1$ is shown below as an example. Let $H'$ be the part of the Hamiltonian $H_{hfc}+H_{B_1}$ acting on the 24 states with $I_2=1$ from Table \ref{tab:ordering10}.
\begin{align}
    H' = a_1 \times U_1 \, \Lambda_1 \, U_1 + a_2 \times U_2 \, \Lambda_2 \, U_2 - \frac{b_1}{2} \times \mathbb{1}_{12} \otimes Z
\end{align}
Using the matrix elements $\lambda_{1}$ and $\lambda_2$ from Table \ref{tab:ordering10}, we have the matrices $\Lambda_1$ and $\Lambda_2$ shown in equation (C2). In this, and in the matrices shown subsequently, the matrix elements not shown should be assumed to be zeros.
\begin{align}
    \Lambda_1 = \mathbb{1}_3 \otimes
    \begin{bmatrix}
    0.5 & & & & & & & \\
    & 0.5 & & & & & & \\
    & & -1 & & & & & \\
    & & & 0.5 & & & & \\
    & & & & -1 & & & \\
    & & & & & 0.5 & & \\
    & & & & & & 0 & \\
    & & & & & & & 0
    \end{bmatrix},
\quad
    \Lambda_2 = 
    \begin{bmatrix}
    0.5 & & & & &\\
    & 0.5 & & & & \\
    & & \underbrace{\ddots}_{\times 3}& & &\\
    & & & -1 & & \\
    & & & & 0.5 &\\
    & & & & & \underbrace{\ddots}_{\times 7}\\
    \end{bmatrix}.
\end{align}
Now let $CG_i$ denote a Clebsch-Gordan table of spin addition $i \oplus 1/2$ as follows. 
\begin{align}
CG_1 = 
\begin{bmatrix}
1 & & & & & \\
& \sqrt{\frac{1}{3}} & \sqrt{\frac{2}{3}} & & & \\
& \sqrt{\frac{2}{3}} & -\sqrt{\frac{1}{3}} & & & \\
& & & \sqrt{\frac{2}{3}} & \sqrt{\frac{1}{3}} & \\
& & & \sqrt{\frac{1}{3}} & -\sqrt{\frac{2}{3}} & \\
& & & & & 1
\end{bmatrix}, 
\quad
CG_0 = 
\begin{bmatrix}
1 & \\
& 1
\end{bmatrix}
\end{align}

Due to our choice of state ordering, $U_1$ from equation (C1) can be set as 
$
U_1 = \mathbb{1}_3 \otimes
\begin{bmatrix}
CG_1 & \\
& CG_0
\end{bmatrix}
$
In contrast, $U_2$ is more complex, because it is composed from spin addition corresponding to the same $\ket{I_1, m_1}$. In general, there is no compact mathematical expression for $U_2$, because it is the unitary matrix that mixes the electronic state with the nuclear states $\ket{I_2,m_2}$ as seen in Table \ref{tab:ordering10}. Therefore, it is best described by the pseudocode shown in Procedure \ref{alg:u2}.

\begin{table}[H]
    \begin{subtable}[h]{0.3\textwidth}
        \centering
        \begin{tabular}{ |p{1.1cm}|p{1.1cm}|p{0.6cm}|p{0.8cm}|p{0.8cm}| }
    \hline
    \textbf{$\ket{I_2,m_2}$} & \textbf{$\ket{I_1,m_1}$} & \textbf{$s_1$} & $\lambda_1$ & $\lambda_2$\\
    \hline
    \hline
    $\ket{6,6}$ & $\ket{1,1}$ & $\ket{\uparrow}$ & 0.5 & 3\\
    & & $\ket{\downarrow}$ & 0.5 & 3\\
    & $\ket{1,0}$ & $\ket{\uparrow}$ & -1 & 3\\ 
    & & $\ket{\downarrow}$ & 0.5 & 3\\
    & $\ket{1,-1}$ & $\ket{\uparrow}$ & -1 & 3\\
    & & $\ket{\downarrow}$ & 0.5 & 3\\
    & $\ket{0,0}$ & $\ket{\uparrow}$ & 0 & 3\\ 
    & & $\ket{\downarrow}$ & 0 & 3\\
    \hline
    $\ket{6,5}$ & $\ket{1,1}$ & $\ket{\uparrow}$ & 0.5 & -3.5\\
    & & $\ket{\downarrow}$ & 0.5 & 3\\
    & $\ket{1,0}$ & $\ket{\uparrow}$ & -1 & -3.5\\ 
    & & $\ket{\downarrow}$ & 0.5 & 3\\
    & $\ket{1,-1}$ & $\ket{\uparrow}$  & -1 & -3.5\\
    & & $\ket{\downarrow}$ & 0.5 & 3\\
    & $\ket{0,0}$ & $\ket{\uparrow}$ & 0 & -3.5\\ 
    & & $\ket{\downarrow}$ & 0 & 3\\
    \hline
    $\vdots$ & & & & \\
    \hline
    $\ket{6,-6}$ & $\ket{1,1}$ & $\ket{\uparrow}$ & 0.5 & -3.5\\
    & & $\ket{\downarrow}$ & 0.5 & 3\\
    & $\ket{1,0}$ & $\ket{\uparrow}$ & -1 & -3.5\\ 
    & & $\ket{\downarrow}$ & 0.5 & 3\\
    & $\ket{1,-1}$ & $\ket{\uparrow}$ & -1 & -3.5\\
    & & $\ket{\downarrow}$ & 0.5 & 3\\
    & $\ket{0,0}$ & $\ket{\uparrow}$ & 0 & -3.5\\ 
    & & $\ket{\downarrow}$ & 0 & 3\\
    \hline
    \end{tabular}
        \caption{$I_2=6$}
        \label{tab:ordering6}
    \end{subtable}
    \hfill
    \begin{subtable}[h]{0.3\textwidth}
        \centering
        \begin{tabular}{ |p{1.1cm}|p{1.1cm}|p{0.6cm}|p{0.8cm}|p{0.8cm}| }
    \hline
    \textbf{$\ket{I_2,m_2}$} & \textbf{$\ket{I_1,m_1}$} & \textbf{$s_1$} & $\lambda_1$ & $\lambda_2$\\
    \hline
    \hline
    $\ket{5,5}$ & $\ket{1,1}$ & $\ket{\uparrow}$ & 0.5 & 2.5 \\
    & & $\ket{\downarrow}$ & 0.5 & 2.5\\
    & $\ket{1,0}$ & $\ket{\uparrow}$ & -1 & 2.5\\ 
    & & $\ket{\downarrow}$ & 0.5 & 2.5\\
    & $\ket{1,-1}$ & $\ket{\uparrow}$ & -1 & 2.5\\
    & & $\ket{\downarrow}$ & 0.5 & 2.5\\
    & $\ket{0,0}$ & $\ket{\uparrow}$ & 0 & 2.5\\ 
    & & $\ket{\downarrow}$ & 0 & 2.5\\
    \hline
    $\ket{5,4}$ & $\ket{1,1}$ & $\ket{\uparrow}$ & 0.5 & -3\\
    & & $\ket{\downarrow}$ & 0.5 & 2.5\\
    & $\ket{1,0}$ & $\ket{\uparrow}$ & -1 & -3\\ 
    & & $\ket{\downarrow}$ & 0.5 & 2.5\\
    & $\ket{1,-1}$ & $\ket{\uparrow}$ & -1 & -3\\
    & & $\ket{\downarrow}$ & 0.5 & 2.5\\
    & $\ket{0,0}$ & $\ket{\uparrow}$ & 0 & -3\\ 
    & & $\ket{\downarrow}$ & 0 & 2.5\\
    \hline
    $\vdots$ & & & &  \\
    \hline
    $\ket{5,-5}$ & $\ket{1,1}$ & $\ket{\uparrow}$ & 0.5 & -3\\
    & & $\ket{\downarrow}$ & 0.5 & 2.5\\
    & $\ket{1,0}$ & $\ket{\uparrow}$ & -1 & -3\\ 
    & & $\ket{\downarrow}$ & 0.5 & 2.5\\
    & $\ket{1,-1}$ & $\ket{\uparrow}$ & -1 & -3\\
    & & $\ket{\downarrow}$ & 0.5 & 2.5\\
    & $\ket{0,0}$ & $\ket{\uparrow}$ & 0 & -3\\ 
    & & $\ket{\downarrow}$ & 0 & 2.5\\
    \hline
    \end{tabular}
        \caption{$I_2=5$}
        \label{tab:ordering5}
     \end{subtable}
     \hfill
    \begin{subtable}[h]{0.3\textwidth}
        \centering
        \begin{tabular}{ |p{1.1cm}|p{1.1cm}|p{0.6cm}|p{0.8cm}|p{0.8cm}| }
    \hline
    \textbf{$\ket{I_2,m_2}$} & \textbf{$\ket{I_1,m_1}$} & \textbf{$s_1$} & $\lambda_1$ & $\lambda_2$\\
    \hline
    \hline
    $\ket{4,4}$ & $\ket{1,1}$ & $\ket{\uparrow}$ & 0.5 & 2\\
    & & $\ket{\downarrow}$ & 0.5 & 2\\
    & $\ket{1,0}$ & $\ket{\uparrow}$ & -1 & 2\\ 
    & & $\ket{\downarrow}$ & 0.5 & 2\\
    & $\ket{1,-1}$ & $\ket{\uparrow}$ & -1 & 2\\
    & & $\ket{\downarrow}$ & 0.5 & 2\\
    & $\ket{0,0}$ & $\ket{\uparrow}$ & 0 & 2\\ 
    & & $\ket{\downarrow}$ & 0 & 2\\
    \hline
    $\ket{4,3}$ & $\ket{1,1}$ & $\ket{\uparrow}$ & 0.5 & -2.5\\
    & & $\ket{\downarrow}$ & 0.5 & 2\\
    & $\ket{1,0}$ & $\ket{\uparrow}$ & -1 & -2.5\\ 
    & & $\ket{\downarrow}$ & 0.5 & 2\\
    & $\ket{1,-1}$ & $\ket{\uparrow}$  & -1 & -2.5\\
    & & $\ket{\downarrow}$ & 0.5 & 2\\
    & $\ket{0,0}$ & $\ket{\uparrow}$ & 0 & -2.5\\ 
    & & $\ket{\downarrow}$ & 0 & 2\\
    \hline
    $\vdots$ & & & & \\
    \hline
    $\ket{4,-4}$ & $\ket{1,1}$ & $\ket{\uparrow}$ & 0.5 & -2.5\\
    & & $\ket{\downarrow}$ & 0.5 & 2\\
    & $\ket{1,0}$ & $\ket{\uparrow}$ & -1 & -2.5\\ 
    & & $\ket{\downarrow}$ & 0.5 & 2\\
    & $\ket{1,-1}$ & $\ket{\uparrow}$ & -1 & -2.5\\
    & & $\ket{\downarrow}$ & 0.5 & 2\\
    & $\ket{0,0}$ & $\ket{\uparrow}$ & 0 & -2.5\\ 
    & & $\ket{\downarrow}$ & 0 & 2\\
    \hline
    \end{tabular}
        \caption{$I_2=4$}
        \label{tab:ordering4}
     \end{subtable}
    \begin{subtable}[h]{0.3\textwidth}
        \centering
        \begin{tabular}{ |p{1.1cm}|p{1.1cm}|p{0.6cm}|p{0.8cm}|p{0.8cm}| }
    \hline
    \textbf{$\ket{I_2,m_2}$} & \textbf{$\ket{I_1,m_1}$} & \textbf{$s_1$} & $\lambda_1$ & $\lambda_2$\\
    \hline
    \hline
    $\ket{3,3}$ & $\ket{1,1}$ & $\ket{\uparrow}$ & 0.5 & 1.5\\
    & & $\ket{\downarrow}$ & 0.5 & 1.5\\
    & $\ket{1,0}$ & $\ket{\uparrow}$ & -1 & 1.5\\ 
    & & $\ket{\downarrow}$ & 0.5 & 1.5\\
    & $\ket{1,-1}$ & $\ket{\uparrow}$ & -1 & 1.5\\
    & & $\ket{\downarrow}$ & 0.5 & 1.5\\
    & $\ket{0,0}$ & $\ket{\uparrow}$ & 0 & 1.5\\ 
    & & $\ket{\downarrow}$ & 0 & 1.5\\
    \hline
    $\ket{3,2}$ & $\ket{1,1}$ & $\ket{\uparrow}$ & 0.5 & -2\\
    & & $\ket{\downarrow}$ & 0.5 & 1.5\\
    & $\ket{1,0}$ & $\ket{\uparrow}$ & -1 & -2\\ 
    & & $\ket{\downarrow}$ & 0.5 & 1.5\\
    & $\ket{1,-1}$ & $\ket{\uparrow}$  & -1 & -2\\
    & & $\ket{\downarrow}$ & 0.5 & 1.5\\
    & $\ket{0,0}$ & $\ket{\uparrow}$ & 0 & -2\\ 
    & & $\ket{\downarrow}$ & 0 & 1.5\\
    \hline
    $\vdots$ & & & & \\
    \hline
    $\ket{3,-3}$ & $\ket{1,1}$ & $\ket{\uparrow}$ & 0.5 & -2\\
    & & $\ket{\downarrow}$ & 0.5 & 1.5\\
    & $\ket{1,0}$ & $\ket{\uparrow}$ & -1 & -2\\ 
    & & $\ket{\downarrow}$ & 0.5 & 1.5\\
    & $\ket{1,-1}$ & $\ket{\uparrow}$ & -1 & -2\\
    & & $\ket{\downarrow}$ & 0.5 & 1.5\\
    & $\ket{0,0}$ & $\ket{\uparrow}$ & 0 & -2\\ 
    & & $\ket{\downarrow}$ & 0 & 1.5\\
    \hline
    \end{tabular}
        \caption{$I_2=3$}
        \label{tab:ordering3}
    \end{subtable}
    \hfill
    \begin{subtable}[h]{0.3\textwidth}
        \centering
        \begin{tabular}{ |p{1.1cm}|p{1.1cm}|p{0.6cm}|p{0.8cm}|p{0.8cm}| }
    \hline
    \textbf{$\ket{I_2,m_2}$} & \textbf{$\ket{I_1,m_1}$} & \textbf{$s_1$} & $\lambda_1$ & $\lambda_2$\\
    \hline
    \hline
    $\ket{2,2}$ & $\ket{1,1}$ & $\ket{\uparrow}$ & 0.5 & 1 \\
    & & $\ket{\downarrow}$ & 0.5 & 1\\
    & $\ket{1,0}$ & $\ket{\uparrow}$ & -1 & 1\\ 
    & & $\ket{\downarrow}$ & 0.5 & 1\\
    & $\ket{1,-1}$ & $\ket{\uparrow}$ & -1 & 1\\
    & & $\ket{\downarrow}$ & 0.5 & 1\\
    & $\ket{0,0}$ & $\ket{\uparrow}$ & 0 & 1\\ 
    & & $\ket{\downarrow}$ & 0 & 1\\
    \hline
    $\ket{2,1}$ & $\ket{1,1}$ & $\ket{\uparrow}$ & 0.5 & -1.5\\
    & & $\ket{\downarrow}$ & 0.5 & 1\\
    & $\ket{1,0}$ & $\ket{\uparrow}$ & -1 & -1.5\\ 
    & & $\ket{\downarrow}$ & 0.5 & 1\\
    & $\ket{1,-1}$ & $\ket{\uparrow}$ & -1 & -1.5\\
    & & $\ket{\downarrow}$ & 0.5 & 1\\
    & $\ket{0,0}$ & $\ket{\uparrow}$ & 0 & -1.5\\ 
    & & $\ket{\downarrow}$ & 0 & 1\\
    \hline
    $\vdots$ & & & &  \\
    \hline
    $\ket{2,-1}$ & $\ket{1,1}$ & $\ket{\uparrow}$ & 0.5 & -1.5\\
    & & $\ket{\downarrow}$ & 0.5 & 1\\
    & $\ket{1,0}$ & $\ket{\uparrow}$ & -1 & -1.5\\ 
    & & $\ket{\downarrow}$ & 0.5 & 1\\
    & $\ket{1,-1}$ & $\ket{\uparrow}$ & -1 & -1.5\\
    & & $\ket{\downarrow}$ & 0.5 & 1\\
    & $\ket{0,0}$ & $\ket{\uparrow}$ & 0 & -1.5\\ 
    & & $\ket{\downarrow}$ & 0 & 1\\
    \hline
    \end{tabular}
        \caption{$I_2=2$}
        \label{tab:ordering2}
     \end{subtable}
     \hfill
    \begin{subtable}[h]{0.3\textwidth}
        \centering
        \begin{tabular}{ |p{1.1cm}|p{1.1cm}|p{0.6cm}|p{0.8cm}|p{0.8cm}| }
    \hline
    \textbf{$\ket{I_2,m_2}$} & \textbf{$\ket{I_1,m_1}$} & \textbf{$s_1$} & $\lambda_1$ & $\lambda_2$\\
    \hline
    \hline
    $\ket{1,1}$ & $\ket{1,1}$ & $\ket{\uparrow}$ & 0.5 & 0.5\\
    & & $\ket{\downarrow}$ & 0.5 & 0.5\\
    & $\ket{1,0}$ & $\ket{\uparrow}$ & -1 & 0.5\\ 
    & & $\ket{\downarrow}$ & 0.5 & 0.5\\
    & $\ket{1,-1}$ & $\ket{\uparrow}$ & -1 & 0.5\\
    & & $\ket{\downarrow}$ & 0.5 & 0.5\\
    & $\ket{0,0}$ & $\ket{\uparrow}$ & 0 & 0.5\\ 
    & & $\ket{\downarrow}$ & 0 & 0.5\\
    \hline
    $\vdots$ & & & & \\
    \hline
    $\ket{1,-1}$ & $\ket{1,1}$ & $\ket{\uparrow}$ & 0.5 & -1\\
    & & $\ket{\downarrow}$ & 0.5 & 0.5\\
    & $\ket{1,0}$ & $\ket{\uparrow}$ & -1 & -1\\ 
    & & $\ket{\downarrow}$ & 0.5 & 0.5\\
    & $\ket{1,-1}$ & $\ket{\uparrow}$  & -1 & -1\\
    & & $\ket{\downarrow}$ & 0.5 & 0.5\\
    & $\ket{0,0}$ & $\ket{\uparrow}$ & 0 & -1\\ 
    & & $\ket{\downarrow}$ & 0 & 0.5\\
    \hline
    $\ket{0,0}$ & $\ket{1,1}$ & $\ket{\uparrow}$ & 0.5 & 0\\
    & & $\ket{\downarrow}$ & 0.5 & 0\\
    & $\ket{1,0}$ & $\ket{\uparrow}$ & -1 & 0\\ 
    & & $\ket{\downarrow}$ & 0.5 & 0\\
    & $\ket{1,-1}$ & $\ket{\uparrow}$ & -1 & 0\\
    & & $\ket{\downarrow}$ & 0.5 & 0\\
    & $\ket{0,0}$ & $\ket{\uparrow}$ & 0 & 0\\ 
    & & $\ket{\downarrow}$ & 0 & 0\\
    \hline
    \end{tabular}
        \caption{$I_2=1,0$}
        \label{tab:ordering10}
     \end{subtable}
     \caption{Basis state ordering for the DMB radical cation. $s_1$ is the electronic spin on the radical cation, and $\lambda_{1,2}$ are the eigenvalues of the two parts of $H_{hfc}$ corresponding to different groups of magnetically equivalent nuclei.}
     \label{tab:ordering_dmb}
\end{table}

\begin{algorithm}[H]
\caption{Setting up $U_{2}$}\label{alg:u2}
\begin{algorithmic}
\For{$x,y$ in zip(vertical nonzero indices of $CG_{1}$, horizontal nonzero indices of $CG_{1}$)}
    \State $x' \gets x + (x//2) \times 6$
    \State $y' \gets y + (y//2) \times 6$
    \State $U_2[x'][y'] \gets CG_1[x][y]$
    \State $U_2[x' + 2][y' + 2] \gets CG_1[x][y]$
    \State $U_2[x' + 4][y' + 4] \gets CG_1[x][y]$
    \State $U_2[x' + 6][y' + 6] \gets CG_1[x][y]$
\EndFor
\end{algorithmic}
\end{algorithm}

Note that $H'$ is of size $24 \times 24$. As mentioned above, all-zero matrices need to be inserted to round up the Hamiltonian size to the nearest power of 
$2$.

\begin{align}
    H = 
    \begin{bmatrix}
    H' & & & \\
    & \mathbb{0}_8 & & \\
    & & H' & \\
    & & & \mathbb{0}_8
    \end{bmatrix}
    + \frac{b_2}{2}
    \begin{bmatrix}
    -\mathbb{1}_{24} & & & \\
    & \mathbb{0}_8 & & \\
    & & \mathbb{1}_{24} & \\
    & & & \mathbb{0}_8
    \end{bmatrix}
\end{align}

We run $H$ on a mixed state circuit as shown in Fig. \ref{fig:circ2} of Appendix \ref{appendix:mixed} and obtain the outcome $S_{I_2}(t)$ for $I_2=1$.
\begin{align}
    S_1(t) = \sum_{n=1}^{12} \frac{s_1^n(t)}{16} + \sum_{m=1}^{4} \frac{1}{16}
\end{align}
In the sum above, $n$ counts the nuclear basis states, and $m$ keeps track of the appended zeros.  After measuring all $S_{I_2}(t)$, we can recover $S(t)$, utilizing Table \ref{tab:addingspins_diffhfc}. In equation (C6), in each parenthetical term, the effect of the appended zeros is subtracted away, and then the correct degeneracy is multiplied.

\begin{align}
    S(t) = &\bigg[ \bigg( S_6(t) - \frac{12}{2^6} \bigg) \cdot 2^6 \cdot \textbf{1}
    + \bigg( S_5(t) - \frac{20}{2^6} \bigg) \cdot 2^6 \cdot \textbf{11} 
    + \bigg( S_4(t) - \frac{28}{2^6} \bigg) \cdot 2^6 \cdot \textbf{54} 
    + \bigg( S_3(t) - \frac{4}{2^5} \bigg) \cdot 2^5 \cdot \textbf{154} \\
    &+ \bigg( S_2(t) - \frac{12}{2^5} \bigg) \cdot 2^5 \cdot \textbf{275} 
    + \bigg( S_1(t) - \frac{4}{2^4} \bigg) \cdot 2^4 \cdot \textbf{297} 
    + \bigg( S_0(t) - \frac{0}{2^2} \bigg) \cdot 2^2 \cdot \textbf{132} \bigg] \; / \; 2^{14} \nonumber
\end{align}

\begingroup
\setlength{\tabcolsep}{6pt} 
\renewcommand{\arraystretch}{1} 
\begin{table}[H]
   \caption{Spin addition to count the total-spin states of the 12 magnetically equivalent protons of DMB. Counts continue from the last row of Table \ref{tab:addingspins}.} 
   \label{tab:addingspins_diffhfc}
   \small
   \centering
   \resizebox{0.8\textwidth}{!}{%
   \begin{tabular}{l|c|c|c|c|c|c|c|c|c|c|c|c|r}
   \toprule
   \backslashbox{\textbf{Number}}{\textbf{Spin}}
    & \textbf{0} & \textbf{$\mathbf{\frac{1}{2}}$} & \textbf{1} & \textbf{$\mathbf{\frac{3}{2}}$} & \textbf{2} & \textbf{$\mathbf{\frac{5}{2}}$} & \textbf{3} & \textbf{$\mathbf{\frac{7}{2}}$} & \textbf{4} & \textbf{$\mathbf{\frac{9}{2}}$} & \textbf{5} & \textbf{$\mathbf{\frac{11}{2}}$} & \textbf{6}\\ 
   \midrule
   9 & & 42 & & 48 & & 27 & & 8 & & 1 & & &\\
   \midrule
   10 & 42 & & 90 & & 75 & & 35 & & 9 & & 1 & &\\
   \midrule
   11 & & 132 & & 165 & & 110 & & 44 & & 10 & & 1 &\\
   \midrule
   12 & \textbf{132} & & \textbf{297} & & \textbf{275} & & \textbf{154} & & \textbf{54} & & \textbf{11} & & \textbf{1}\\
   \bottomrule
   \end{tabular}
   }%
\end{table}
\endgroup

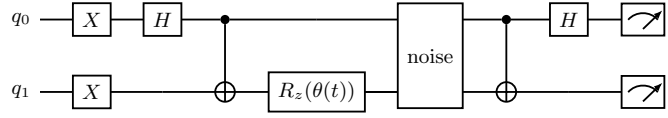
\begin{figure}[H]
    \centering
    \label{fig:dmb_high_circ}
    \caption{Quantum circuit simulating the thermal relaxation of the DMB$^+$/PTP$^-$ radical pair system using the inherent qubit noise. $S(t)$ is obtained using the simulator and then encoded into the $Z$-rotation gate on the hardware. The rotation parameter is $\theta(t) = 2 \cos^{-1} (\sqrt{S(t)})$. Thermal relaxation can be added to the circuit using the Kraus or inherent noise simulation methods.}
    \label{fig:hardware_encode}
    \resizebox{0.5\textwidth}{!}{%
    \begin{quantikz}
    \lstick{$q_0$}& \gate{X} & \gate{H} & \ctrl{1} & \qw & \gate[2]{\textrm{noise}} & \ctrl{1} & \gate{H} & \meter{} \\
    \lstick{$q_1$}& \gate{X} & \qw & \targ{} & \gate{R_z(\theta(t))} & \qw & \targ{} & \qw & \meter{}
    \end{quantikz}
    }%
\end{figure}

The unitary matrices for the DMB$^+$/PTP$^-$ radical pair Hamiltonian evolution are efficiently simulated using Qiskit \texttt{Aer} (Fig. \ref{fig:dmb_ratio}). On a quantum hardware, however, the full Hamiltonian simulation in the presence of the inherent qubit noise is currently not feasible. That is because when generic decomposition methods are used to decompose these unitaries into one and two-qubit gates available on the hardware, the circuit depth exceeds currently implementable limits. Instead, we obtain the coherent time evolution $S(t)$ of the singlet state probability from the simulator, and encode it in a $Z$-rotation gate as shown in Fig. \ref{fig:hardware_encode}. The noise is simulated using either the inherent qubit noise or the Kraus method.

\section{Details of thermal relaxation simulation leveraging inherent qubit decoherence}
\label{appendix:inherent}
To match the decay rates of the quantum computer with that of the experiment, we have to perform several steps. First, we let the set $\{S,T_0,T_+,T_-\}$ denote the undamped, coherent singlet and triplet state measurement probabilities respectively. Next, we simulate the Hamiltonian on the hardware for any nuclear initial state $\ket{k}$ of consideration without accounting for the noise. This simulation is inherently noisy since it is run on a real quantum hardware, and we obtain the corresponding set of probabilities $\{\tilde{S}_k, \tilde{T}_{0,k}, \tilde{T}_{+,k}, \tilde{T}_{-,k}\}$. Finally, we replace the Hamiltonian circuit with a set of identity gates that have the same duration as the Hamiltonian circuit. This allows us to identify the noise that was present during the Hamiltonian simulation. We set the corresponding probabilities $\{S', T_0', T_+', T_-'\}$ and solve the correction equations \cite{rost2020simulation} to recover the undamped statistics $\{S_k,T_{0,k},T_{+,k},T_{-,k}\}$ as follows:

\begin{align}
    &T_{+,k} = (\tilde{T}_{+,k} - T_+') / (1 - 4T_+') \\
    &T_{-,k} = (\tilde{T}_{-,k} - T_-') / (1 - 4T_-')
\end{align}
\begin{align}
    &A_k = \tilde{S}_k - T_{+,k} T_+' - T_{-,k} T_-' \\
    &B_k = \tilde{T}_{0,k} - T_{+,k} T_+' - T_{-,k} T_-'
\end{align}
\begin{align}
   &S_k = (A_k S' - B_k T_0') / (S'^2 - T_0'^2) \\
   &T_{0,k} = (B_k S' - A_k T_0') / (S'^2 - T_0'^2)
\end{align}

To incorporate the desired noise that imitates decoherence in lab experiments, we construct a circuit shown in Fig. \ref{fig:circ10} that simulates the desired $T_1$ and $T_2$ decay of the radical ion pairs. Since the radical ion pairs in the lab experiments decay faster than qubits on the quantum hardware, we append $N$ identity gates as wait cycles. Choosing $N = \frac{T_{qubit}}{T_{RP} t_{identity}}t$ at every time point ($t$) makes the circuit wait long enough so that the qubit decay matches the decay of the radical ion pairs. Note that $T_{qubit}$ denotes the arithmetic average of $T_1$ and $T_2$. 

Additionally, an even number of $X$ gates (aka echo pulses) are inserted in between identity gates to correct for a hardware precession drift over time. Echo pulses flip $\ket{\downarrow}$ to $\ket{\uparrow}$ during half of the runtime allowing the phase accumulation to cancel instead of compound, and also mimic an amplitude damping channel under infinite temperature. 

For the lab experiments conducted at zero field, it was found that $T_1=T_2=T_{RP}$, which matched well with $T_1 \approx T_2$ found in qubits. For high field experiments, we considered an approximation that $T_1 = \infty$ and utilized the Kraus method from Fig. \ref{fig:circ8} without amplitude damping. To that end, we denote a set of probabilities with the desired $T_1$ and $T_2$ decay $\{S'', T_0'', T_+'', T_-'' \}$, and compute the noisy singlet state probabilities $S_k^n$ for each nuclear initial state $\ket{k}$ with the desired noise parameters as shown below:
\begin{align}
    S_k^n = S_k S'' + T_{0,k} T_0'' + T_{+,k} T_+'' + T_{-,k} T_-''
\end{align}

\end{document}